\documentclass[aps,prb,twocolumn,showpacs,floatfix,superscriptaddress]{revtex4-1}
\usepackage{graphicx,color}
\usepackage{amsfonts}
\usepackage[figuresright]{rotating}
\usepackage{amssymb}
\usepackage{amsmath}
\usepackage{psfrag}
\usepackage{subfigure}
\usepackage{multirow}
\usepackage{tabularx}
\usepackage{textcomp}
\usepackage{units}
\usepackage{hyperref}
\usepackage{adjustbox}
\hypersetup{
 pdfnewwindow=true, colorlinks=true,
 linkcolor=blue, anchorcolor=blue,
 citecolor=blue, filecolor=blue,
 menucolor=blue, urlcolor=blue}

\usepackage{graphicx}
\usepackage{dcolumn}
\usepackage{bm}
\usepackage{color}
\makeatletter

\begin{document}
\title{{ Metastable} piezoelectric group IV monochalcogenide monolayers\\
with a buckled honeycomb structure}

\author{Shiva P.\ \surname{Poudel}}
\email{sppoudel@uark.edu}
\affiliation{Department of Physics, University of Arkansas, Fayetteville, AR 72701, USA}

\author{Salvador\ \surname{Barraza-Lopez}}
\email{sbarraza@uark.edu}
\affiliation{Department of Physics, University of Arkansas, Fayetteville, AR 72701, USA}
\affiliation{Institute for Nanoscience and Engineering, University of Arkansas, Fayetteville, Arkansas 72701, USA}

\date{\today} 

\begin{abstract}
{  Multiple two-dimensional materials are being na\"ively termed {\em stable} on the grounds of displaying phonon dispersions with no negative frequencies, and of not collapsing on molecular dynamics calculations at fixed volume. But, if these phases do not possess the smallest possible structural energy, how does one understand and establish their actual {\em meta-}stability? To answer this question,} twelve two-dimensional group-IV monochalcogenide monolayers (SiS, SiSe, SiTe, GeS, GeSe, GeTe, SnS, SnSe, SnTe, PbS, PbSe, and PbTe) with a buckled honeycomb atomistic structure--belonging to symmetry group P3m1--and an out-of-plane intrinsic electric polarization are shown to be metastable by three independendent methods. First, we uncover a coordination-preserving structural transformation from the low-buckled honeycomb structure onto the lower-energy Pnm2$_1$ (or Pmmn for PbS, PbSe, and PbTe) phase to estimate {\em energy barriers} $E_B$ that must be overcome during such  structural transformation. Using the curvature of the local minima and $E_B$ as inputs to Kramers escape formula, large escape times are found, implying the structural metastability of the buckled honeycomb phase { (nevertheless, and with the exception of PbS and PbSe, these phases display escape times ranging from 700 years to multiple times the age of the universe, and can be considered ``stable'' for practical purposes only in that relative sense)}. The second demonstration is provided by phonon dispersion relations that include the effect of long-range Coulomb forces and display no negative vibrational modes. The third and final demonstration of structural metastability is furnished by room-temperature {\em ab initio} molecular dynamics for selected compounds. The magnitude of the electronic band gap evolves with chemical composition. Different from other binary two-dimensional compounds such as transition metal dichalcogenide monolayers and hexagonal boron nitride monolayers which only develop an in-plane piezoelectric response, the twelve group-IV monochalcogenide monolayers with a buckled honeycomb structure also display out-of-plane piezoelectric properties.
\end{abstract}

\maketitle

\section{Introduction}
{\em Stereochemistry} studies the multiple possible atomistic arrangements of chemical compounds. As an incipient example of stereochemical behavior, graphite is the most stable carbon allotrope under standard temperature and pressure conditions. Nevertheless---and despite it not being a structural ground state---diamond is { a metastable} carbon allotrope at room temperature and atmospheric pressure { that can last for over multiple centuries}.

References~\onlinecite{marzari,Haastrup_2018} and others list a vast amount of yet-to-be-experimentally synthesized 2D compounds. Along these lines, theoretical predictions of two-dimensional group-IV monochalcogenides \cite{tritsaris_jap_2013_sns,singh_apl_2014_ges_gese_sns_snse,Hu2016,ref1,Gu2019} are slowly but surely finding experimental confirmation\cite{Chang2016,Higashitarumizu20_NC_SnS,Chang20_arxiv_SnSe} while new and remarkable structures---such as one-dimensional, chiral GeS\cite{sutterNature2019}---are being experimentally found. Besides the well-known two-dimensional phase with a Pnm2$_1$ group symmetry,\cite{kaxiras,singh_apl_2014_ges_gese_sns_snse,rodin_prb_2016_sns} group-IV monochalcogenides have also been predicted to form a 2D phase with Pma2 symmetry { (SiS)},\cite{yang_nanolett_2015_sis} a so-called $A-$MX phase (whose symmetry group was not identified),\cite{Hu2016} and a buckled honeycomb phase.\cite{Hu2016,ref1,GeTe,Gu2019}

{ We make a case for the lack of discussions of structural metastability, by revising the language employed when introducing three extremely popular two-dimensional materials with a low-buckled honeycomb structure. Afterwards, we facilitate processes that can be employed to understand metastability, which should be addressed as new 2D materials are laid out.}

{ Indeed, when introducing blue phosphorene (a low-buckled honeycomb two-dimensional form of phosphorus), the following is said concerning structural stability: ``blue and black phosphorus are {\em equally  stable};'' ``the relative energy with respect to the black phosphorus structure illustrate ... that the blue phosphorus structure is equally {\em stable}.'' ``We find blue phosphorus to be {\em nearly as stable} as black phosphorus, the most stable phosphorus allotrope.''\cite{Zhu2014} Germanene and stanene have a nine-fold coordination in their most stable two-dimensional phases\cite{Rivero2014}. Low-buckled phases, on the other hand, are three-fold coordinated. Nevertheless, it has been said that ``germanium can have {\em stable}, two-dimensional, low-buckled, honeycomb structures,'' that ``the {\em stability} of low-buckled structures of ... Ge are further tested by extensive {\em ab initio} finite temperature molecular dynamics calculations,'' and that ``the present analysis together with calculated phonon dispersion curves provides a {\em stringent test for the stability of LB honeycomb structure of ... Ge}.''\cite{Cahangirov2009} Similarly, one reads that ``a low-buckled configuration is found to be {\em more stable} for stanene.''\cite{PhysRevLett.111.136804} In what follows, we focus on diamond and on the low-buckled phases of group IV monochalcogenide monolayers with a buckled honeycomb structure, and demonstrate processes to understand the relative stability of phases located at local minima in the elastic energy landscape. We then provide a study of the electronic, elastic, and piezoelectric properties of the latter compounds.}

GeS, GeSe, SnS, and SnSe monolayers with a buckled honeycomb structure were studied in Ref.~\onlinecite{Hu2016}. In that work, {\em ab initio} molecular dynamics (AIMD) within the NVT ensemble (an ensemble with a constant number of atoms $N$, constant volume $V$,  and constant temperature $T$) were performed for up to 10 ps at room temperature. Vibrational frequencies at the $\Gamma-$point were also calculated, to find no vibrational modes with imaginary frequencies. Binding energies were reported, as well as electronic structures that show these materials to be semiconductors with indirect electronic band gaps. The out-of-plane intrinsic electric polarization, phonon dispersions, and AIMD calculations using the NVT ensemble were reported in Ref.~\onlinecite{ref1} for GeS and GeSe.

Duerloo and coworkers calculated the piezoelectric properties of hexagonal boron nitride (hBN) monolayers and of transition metal dichalcogenide (2H-MoS$_2$, 2H-MoSe$_2$, 2H-MoTe$_2$, 2H-WS$_2$, 2H-WSe$_2$, and 2H-WTe$_2$) monolayers,\cite{duerloo} and piezoelectricity has been experimentally confirmed for MoS$_2$ and hBN monolayers since.\cite{piezoMoS2,piezohBN} {Unlike the group-IV monochalcogenide monolayers with a buckled honeycomb structure studied here}, these 2D materials are mirror-symmetric with respect to the plane defined by the hBN monolayer, or by the plane defined by Mo or Te atoms in the 2H-dichalcogenide monolayers, and hence do not develop an out-of-plane piezoelectric response.

Using Duerloo and coworkers' procedure,\cite{duerloo} the {\em in-plane piezoelectric coefficients} were reported for GeS, GeSe, SnS, and SnSe monolayers with a buckled honeycomb structure.\cite{Hu2016} Nevertheless, inversion symmetry is broken in this 2D phase, and we will report the resulting out-of-plane intrinsic electric polarization and the out-of-plane piezoelectric response that is missing in Ref.~\onlinecite{Hu2016}. The existence of (i) an out-of-plane intrinsic polarization and of (ii) a tunable electronic bandgap by material thickness\cite{ref1} or by in-plane strain makes honeycomb buckled group-IV monochalcogenide monolayers\cite{Gu2019} relevant materials for water-splitting applications.\cite{ref1,GeTe,Gu2019} As it will be shown here, further band gap tunability may be achieved {\em via} chemical composition when the twelve possible compounds in this family are considered (only five such compounds---GeS, GeSe, SnS, SnSe, and GeTe---have been studied thus far).

The manuscript is structured as follows: Computational methods are disclosed in Sec.~\ref{sec:cm}. Then, the concepts of the energy barrier $E_B$ separating two structural phases, and of Kramers escape time are exemplified in a diamond-to-graphite transformation in Sec.~\ref{diamond}. The group symmetries of group-IV monochalcogenide monolayers with a buckled honeycomb structure are determined in Sec.~\ref{structure}, and their structural metastability at room temperature is established in Sec.~\ref{stability} using three complementary approaches, including: (i) The determination of an energy path joining the high-energy P3m1 phase to the low-energy Pnm2$_1$ (or Pmmn) structure; the path shows tall energy barriers, and escape times are obtained using Kramers escape formula.\cite{toledo} (ii) Phonon dispersion calculations---in which the long-range effective charge contribution to the dynamical matrix\cite{PhysRevB.1.910} is included---displaying no negative frequencies. (iii) AIMD calculations within the NPT ensemble at room temperature for two such compounds that show small variations in interatomic distances from the reference, zero-temperature structure. Chemical composition is shown to be an additional handle to engineer the magnitude of the electronic band gap. This work ends with a revision of the piezoelectric properties of these materials in Sec.~\ref{piezo}, including their out-of-plane piezoelectric response. Conclusions are provided in Sec.~\ref{conclusions}.

\section{Computational methods}\label{sec:cm}
\emph{Ab initio} calculations were performed with the \textit{VASP} code\cite{vasp,vasp1} on freestanding monochalcogenide monolayers with a buckled honeycomb structure. We employed a 18$\times$18$\times$1 $k-$point grid and a cutoff energy of 500 eV. The energy and force convergence criteria were set to $10^{-6}$ eV and $10^{-3}$ eV/\AA{}, respectively. We employ exchange-correlation functionals that include self-consistent van der Waals corrections\cite{reviewvdw} with the optPBE-vdW functional.\cite{klimes1,klimes2,klimes3} Dipole corrections were employed unless explicitly indicated.

Phonon dispersion calculations were performed with \emph{PHONOPY}\cite{phonophy} and \emph{VASP} on a 7$\times$7$\times$1 supercell. Finite atomic displacements were set at 0.005 \AA. In these supercell calculations, the $k-$point grid was set to 5$\times$5$\times$1 and the cutoff energy  remained at 500 eV. The energy convergence remained at $10^{-6}$ eV as well. Importantly, the effect of Born charges was included in the phonon dispersion calculations in order to properly describe their long-wavelength behavior.\cite{PhysRevB.1.910} The magnitude of the polarization vector $\mathbf{P}$ was obtained using the modern theory of polarization.\cite{modern}

Additionally, we performed room-temperature AIMD calculations \cite{Car-Parrinello} with the \emph{SIESTA} code\cite{siesta} without dipole corrections for two representative SiS and PbS monolayers for 15 ps on a rectangular supercell containing 336 atoms, and AIMD calculations for a PbS monolayer at 2,000 K for 2 ps.  This computer code employs localized numeric atomic orbitals (NAOs)\cite{Junquera2001} and norm-conserving Troullier-Martins pseudopotentials \cite{Troullier} tuned in-house\cite{rivero} with van der Waals corrections of the Berland-Per Hyldgaard (BH) type\cite{Hyldgaard} as implemented by Rom\'an-P\'erez and Soler.\cite{soler}  Not needing to fill vacuum with plane waves, the use of NAOs makes AIMD calculations more economic; hence this choice of code for these calculations. The AIMD calculations were performed using the NPT ensemble (constant number of electrons $N$, and at a target ambient pressure $P$ at selected temperatures $T$) using methods described elsewhere.\cite{Mehboudi-nl,Mehboudi-prl,Salvador2018-prb}

Periodic images along the direction perpendicular to the 2D material were separated by a distance $a_3=20$ \AA{} in all calculations involving 2D materials.

\section{Structural metastability of diamond}\label{diamond}

We begin by reminding the reader that atomistic configurations having higher energy than ground-state structures are metastable. A classical example is presented by diamond and graphite, well-known carbon allotropes available under ordinary temperature and pressure conditions.

\begin{figure*}[tb]
\includegraphics[width=0.96\textwidth]{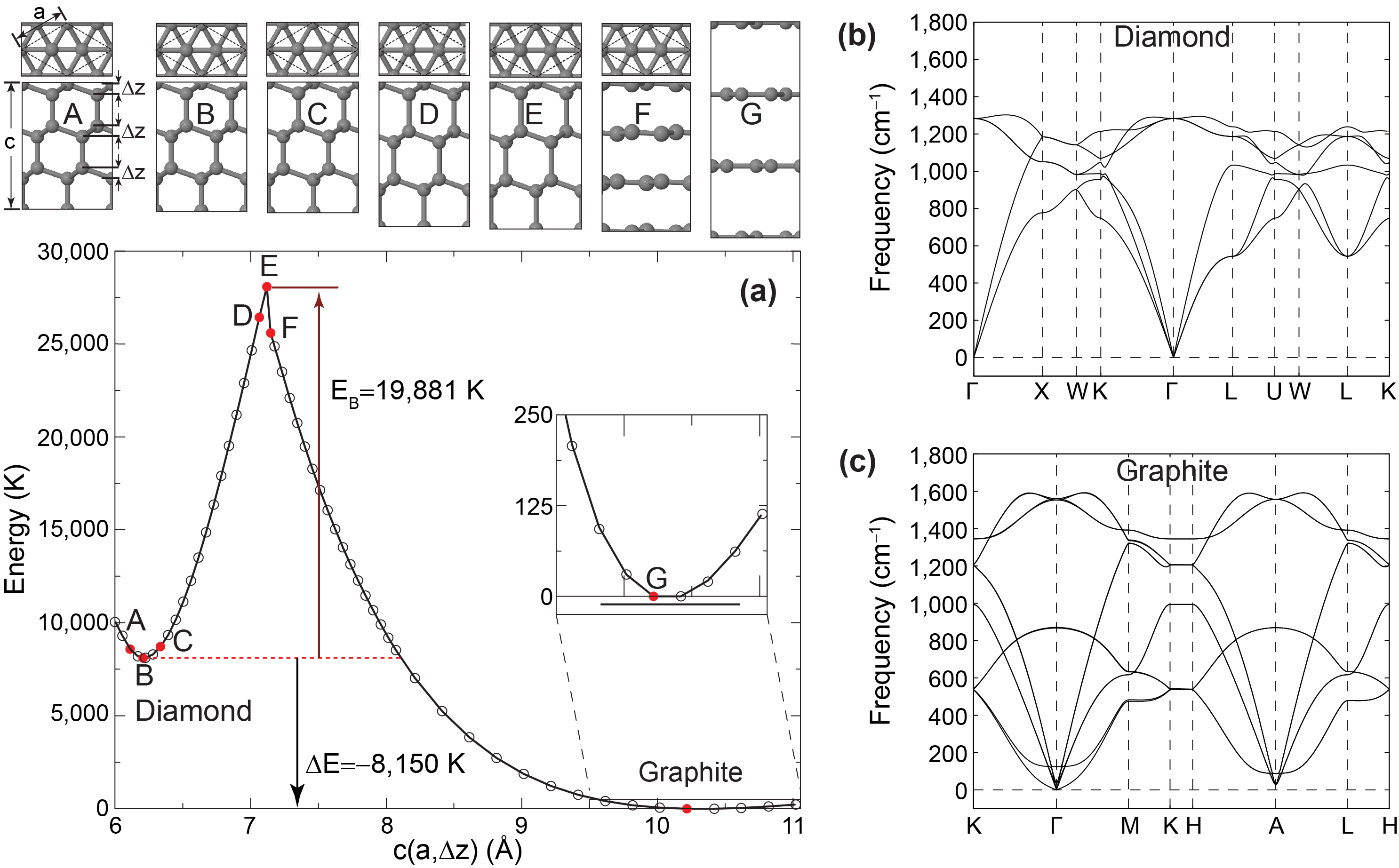}
\caption{Computational demonstration of the metastability of diamond. (a) When the $z-$axis is oriented along the (111) crystallographic direction, diamond can be turned into ABC graphite by following the lowest-energy structural path depicted by the structures shown above. (The structural energy of Bernal graphite is seen as a horizontal straight line at the inset.) Despite of having an energy 8,150 K higher than ABC graphite, it takes an astronomical time of $10^{9}$ years for diamond to tunnel along the horizontal dashed line through the energy barrier $E_B$. (b) The large barrier height guarantees the existence of phonon dispersions with no imaginary frequencies for diamond. (c) Phonon dispersion for the ground-state carbon allotrope, Bernal graphite.}
\label{fig:figure1}
\end{figure*}

Figure \ref{fig:figure1}(a) describes a structural transformation whereby diamond turns into ABC graphite. The leftmost structural diagram atop Fig.~\ref{fig:figure1}(a) contains side and top views of diamond. There, the $z-$axis is oriented along the (111) crystallographic direction on a cell containing six atoms; the cell is characterized by lattice parameters $a$ and $c$, and atomic positions are given by ($0,0,0$), ($a/\sqrt{3},0,\Delta z$), ($a/\sqrt{3},0,c/3$), ($2a/\sqrt{3},0,c/3+\Delta z$), ($2a/\sqrt{3},0,2c/3$), and ($0,0,2c/3+\Delta z$). The point is that the structure can be fully characterized by three independent variables ($a$, $c$, and $\Delta z$). Being nonpolar compounds, no dipole corrections are necessary at this stage.

One may employ the nudged elastic band method\cite{nudged1,nudged2} to estimate the energy barrier in between diamond and graphite. Nevertheless, such method may overestimate the energy barrier (see, {\em e.g.,} Ref.~\onlinecite{tomanek}). The alternative approach followed here is a full sampling of the ($a$,$c$,$\Delta z$) space over a sufficiently wide parameter range: being explicit, $a$ was sampled from 2.38 to 2.53 \AA{} in fifteen steps, $c$ ran from  6.0 to 11 \AA{} in fifty six steps, and $\Delta z$ varied from 0.0 to 0.6 \AA{} in fifteen steps, for a total of 12,375 individual calculations. Any energy value shown in Fig.~\ref{fig:figure1}(a) is acquired on the optimal structure for a given choice of $c$, and Fig.~\ref{fig:figure1}(a) depicts the lowest possible energy barriers in between diamond and graphite. The additional structural plots (B to G) in Fig.~\ref{fig:figure1}(a) illustrate the structural evolution of the six-atom cell into ABC graphite (G) as $c$ increases; capital letters match structures seen on top to the points shown in red along the energy {\em versus} $c$ path in the Figure. Na\"ively, the lower energy $\Delta E=-8,150$ K of graphite with respect to diamond may imply that diamond is not stable. Nevertheless, a large $E_B$ of 19,881 K is the key energetic variable guaranteeing diamond's structural metastability.

The inset in Fig.~\ref{fig:figure1}(a) depicts a horizontal line that represents the---comparatively tiny---energy difference between ABC and Bernal graphite (the energy of Bernal graphite---whose unit cell contains four atoms---has been multiplied by 3/2 for a direct comparison). A standard construct in the theory of glasses---{\em i.e.}, materials that display {\em multiple local minima}---is Kramers escape formula,\cite{toledo} which permits estimating escape times $\tau$ from a metastable minimum onto the ground state structure {qualitatively}; $\tau$ is proportional to the classical oscillation frequency $\omega=\sqrt{\frac{k}{m}}$ at the metastable local minimum centered at point $B$, to the barrier height $E_B$, and to $T$, through the following relation:\cite{rmp1990,toledo}
\begin{equation}\label{eq:escape}
\tau(T) = \lambda\frac{2\pi}{\omega}\exp\left(\frac{E_B}{T}\right),
\end{equation}
with $E_B$ given in $K$, as it is the case in Fig.~\ref{fig:figure1}(a), and $\lambda$ being a dimensionless prefactor. We will consider the {predominant} contribution of the exponential to the escape time, and will assume $\lambda$ to be of the order of unity. The parameter $k$ is obtained by fitting the energy around the local minimum (point $B$) $E=\frac{k}{2}(c-c_D)^2$, where $c_D=6.21$ \AA{} is the magnitude of $c$ for the diamond supercell, and $m$ is the mass of the 6-atom supercell. The frequency $\omega$ turns out to be 43.4 THz, and the escape time is estimated to be $10^9$ years at room temperature. Similar estimates based on the WKB approximation\cite{elastic} yield an even larger $\tau=10^{11}$ years. Given that the age of the universe is roughly $10^{10}$ years, such astronomical escape times provide additional rationale for the apparent stability of diamond, which is additionally confirmed by the lack of negative/imaginary frequencies in Fig.~\ref{fig:figure1}(b). Fig.~\ref{fig:figure1}(c) shows phonon dispersions for ground-state Bernal graphite which, expectedly, display no negative/imaginary frequencies either.

\section{Structure and crystal symmetries of group-IV monochalcogenide monolayers with a buckled honeycomb structure}\label{structure}

\begin{figure}[tb]
\includegraphics[width=0.48\textwidth]{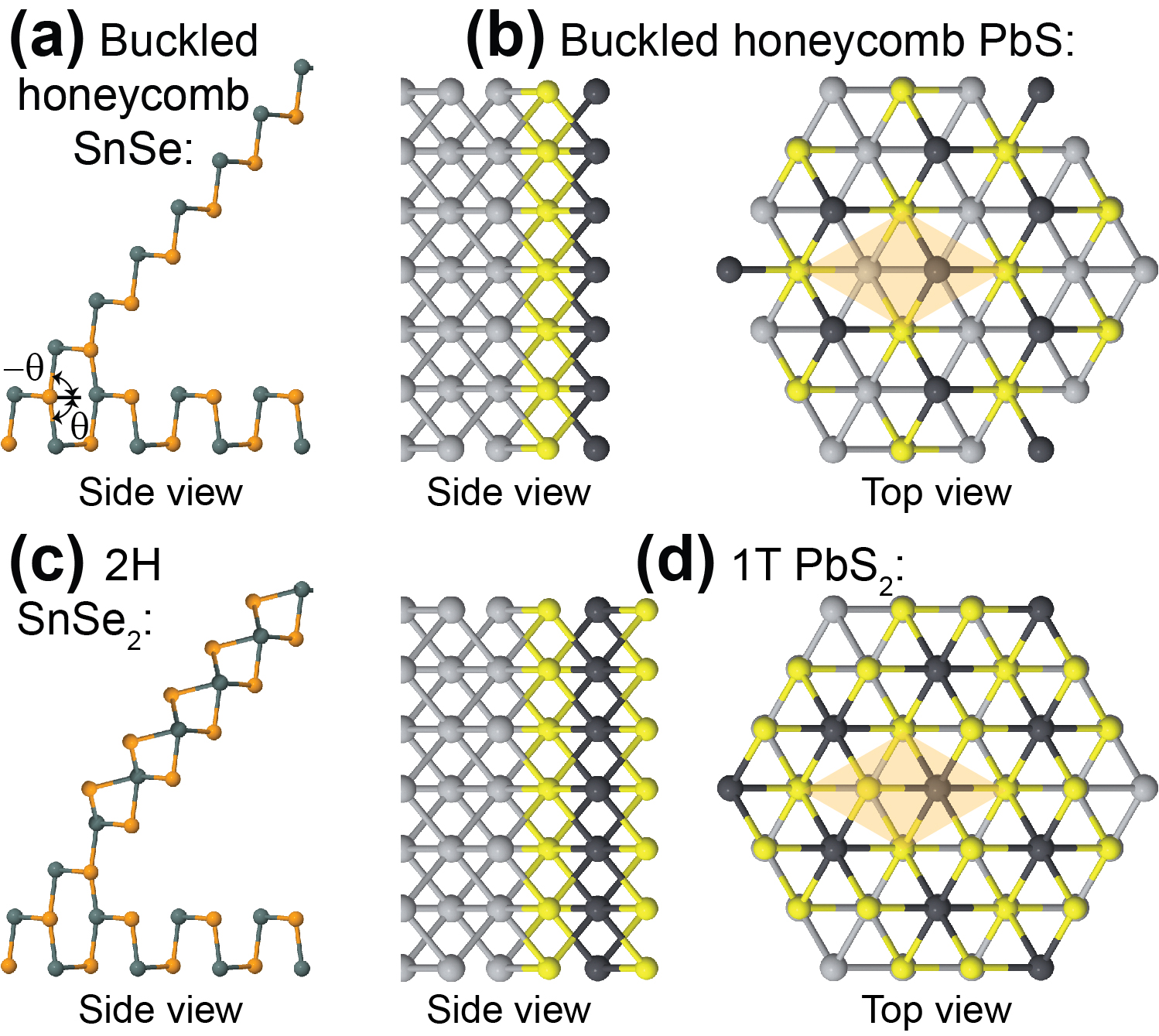}
\caption{Conceptual rendering of group-IV monochalcogenide monolayers with a buckled honeycomb structure: (a) Layered group-IV monochalcogenides (SiS, SiSe, GeS, GeSe, SnS, and SnSe) on the unit cell with Pnm2$_1$ symmetry (horizontal structure) and on a buckled honeycomb structure reminiscent of blue phosphorene. (b) Rhombic and cubic group-IV monochalcogenides (SiTe, GeTe, SnTe, PbS, PbSe, and PbTe) grown along the (111) direction can also give rise to buckled honeycomb structures if grown on an appropriate substrate. (c and d) Additional possible structures could be created, too. The unit cell area is highlighted within orange diamonds in subplots (b) and (d).}
\label{fig:F1N}
\end{figure}

Fig.~\ref{fig:F1N}(a) depicts a cis-to-trans reconstruction similar to the one outlined in Ref.~\onlinecite{Zhu2014} whereby blue phosphorene (two-dimensional phosphorus with a buckled honeycomb structure) was argued for. This transformation may be relevant to create two-dimensional buckled honeycomb phases out of layered group-IV monochalcogenides (SiS, SiSe, GeS, GeSe, SnS, and SnSe). Two-dimensional buckled honeycomb compounds could be grown along the (111) directions on materials with rhombic or cubic bulk structures (SiTe, GeTe, SnTe, PbS, PbSe, and PbTe)\cite{Littlewood1} on suitable commensurate substrates. Growth techniques advance at a fast pace, and two-dimensional films can be created by growth and subsequent wet chemistry.\cite{Nature2019} Fig.~\ref{fig:F1N}(c) shows another possible structure which may compete with the buckled honeycomb one: a 2H dichalcogenide, which has been reported experimentally\cite{ref18,SnS2_2,SnS2_3} and theoretically\cite{Oleynik} already. Similarly, Fig.~\ref{fig:F1N}(d) illustrates the creation of a 1T dichalcogenide by the subsequent growth of a chalcogen layer on the film grown along the (111) direction of monochalcogenides with a non-layered bulk structure. We acknowledge that the buckled honeycomb structures have not been observed in experiment, and the challenges that their realization entails.

\begin{figure}[tb]
\includegraphics[width=0.48\textwidth]{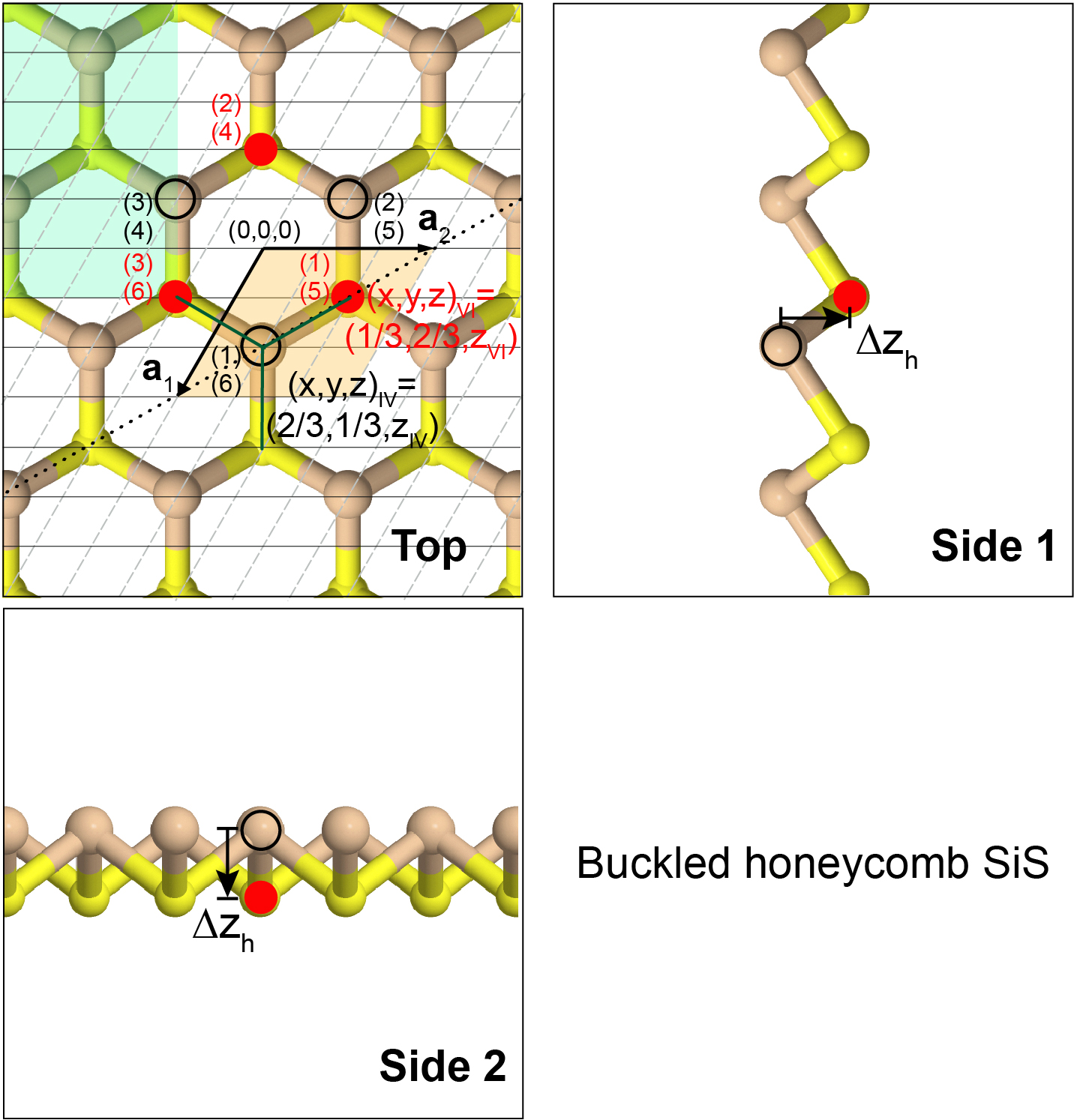}
\caption{\label{fig:fN3} Structure and symmetry operations of buckled honeycomb group-IV monochalcogenide monolayers. The top view lists lattice vectors, the unit cell area (in orange), atomic positions, symmetry operations, a three-fold rotation axis (in dark green), and a mirror plane (dotted black line). Atoms belonging to the $\mathcal{A}$ ($\mathcal{B}$) sublattice are highlighted with open black (full red) circles. A rectangular cell containing four atoms is drawn in light green at the upper left. $\Delta z_h$ is depicted in the side views, which lack inversion symmetry.}
\end{figure}

The group symmetry of buckled honeycomb group-IV monochalcogenides is derived here following a process that starts with graphene's symmetry operations (symmetry group p6mm). According to the International Tables for Crystallography, and as shown in Fig.~\ref{fig:fN3}, the standard choice for lattice vectors is one in which $\mathbf{a}_1=a\left(-\frac{1}{2},-\frac{\sqrt{3}}{2},0\right)$ and $\mathbf{a}_2=a\left(1,0,0\right)$.\cite{Hahn} In standard Crystallography notation, atomic coordinates are expressed in {\em direct} format (cartesian positions of atoms are given by $x\mathbf{a}_1+y\mathbf{a}_2$ in 2D, or $x\mathbf{a}_1+y\mathbf{a}_2+z\mathbf{a}_3$ in 3D). The twelve symmetry operators for graphene (space group 17) have the following coordinates: $x,y$ (symmetry operation 1); $\bar{y},x-y$ (2);  $\bar{x}+y,\bar{x}$ (3); $\bar{x},\bar{y}$ (4); $y,\bar{x}+y$  (5); $x-y,x$ (6);
$\bar{y},\bar{x}$ (7); $\bar{x}+y,y$ (8);  $x,x-y$ (9); $y,x$ (10); $x-y,\bar{y}$ (11); and $\bar{x},\bar{x}+y$ (12).\cite{Hahn} (Here, $\bar{x}=-x$ and $\bar{y}=-y$.\cite{Hahn}) Taking $x=2/3$ and $y=1/3$, symmetry operations (1), (2), (3), (7), (8) and (9) leave atoms within the same sublattice (say, $\mathcal{A}$), while atoms in the $\mathcal{B}$ sublattice are reached through symmetry operations (4), (5), (6), (10), (11) and (12) from an atom originally belonging to the $\mathcal{A}-$sublattice.

Turning into a hBN monolayer, the six symmetry operations above [(4), (5), (6), (10), (11), and (12)] taking a boron atom ($\mathcal{A}-$sublattice) and landing it into a nitrogen atom ($\mathcal{B}-$sublattice) are forbidden. The six allowed symmetry operations are $x,y$; $\bar{y},x-y$;  $\bar{x}+y,\bar{x}$; $\bar{y},\bar{x}$;    $\bar{x}+y,y$; and $x,x-y$; which correspond to symmetry group p3m1.\cite{Hahn} Coordinates $x_\mathcal{A}=2/3$, and $y_\mathcal{A}=1/3$ ($x_\mathcal{B}=1/3$, and $y_\mathcal{B}=2/3$) give atomic positions of the $\mathcal{A}$ ($\mathcal{B}$) sublattice.

The main {structural} difference between a hBN monolayer and buckled honeycomb group-IV monochalcogenide monolayers is the out-of-plane buckling in the latter, which requires supplementing the 2D coordinates of the hBN monolayer with a relative height. We write $x_\mathcal{A}=2/3$, $y_\mathcal{A}=1/3$, $z_\mathcal{A}=z_{IV}/a_3$, and $x_\mathcal{B}=1/3$, $y_\mathcal{B}=2/3$, $z_\mathcal{B}=z_{VI}/a_3$ with $z_{IV}$ and $z_{VI}$ atomic heights in \AA{} such that $\Delta z=a_3(z_\mathcal{B}-z_\mathcal{A})$. The allowed symmetry operations turn out to be: $x,y,z$ (1); $\bar{y},x-y,z$ (2); $\bar{x}+y,\bar{x},z$ (3); $\bar{y},\bar{x},z$ (4); $\bar{x}+y,y,z$  (5); and $x,x-y,z$ (6), corresponding to symmetry group 156 (P3m1, or $C_{3v}^1$).\cite{Hahn}

Group P3m1 has a three-fold rotational symmetry (shown by green axial lines in Fig.~\ref{fig:fN3}, top view), a mirror symmetry shown by a black dotted line, and it lacks an out-of-plane inversion symmetry, as clearly observed in both side views on that Figure. Symmetry operations (1) through (6) were explicitly applied to a SiS monolayer in Fig.~\ref{fig:fN3}, top view, and Table \ref{ta:structure} lists the lattice constant $a_h$ and a (signed) buckling height $\Delta z_h$ of the optimal buckled honeycomb (h) structures. The lack of inversion symmetry confers these materials with an intrinsic out-of-plane electric intrinsic polarization $P_{h,3}$\cite{ref1} (also listed in Table \ref{ta:structure}), and an out-of-plane piezoelectric coefficient that is missing in Ref.~\onlinecite{Hu2016}. Lattice constants are comparable to these reported previously, with differences arising from the different choice of exchange-correlation functional (vdW here, and PBE\cite{PBE} in previous reports; see Table \ref{ta:structure}). The magnitude of the lattice constant on a planar structure, $a_p$ is also listed in Table~\ref{ta:structure} for reasons that will become apparent latter. The average atomic number $\bar{Z}=\frac{Z_{IVA}+Z_{VIA}}{2}$ [with $Z_{IVA}$ ($Z_{VIA}$) the atomic number of the group-IVA (VIA) atom] is listed in subsequent Figures and Tables, to emphasize structural trends in this family of compounds that depend on that variable,\cite{Mehboudi-nl,Shiva} such as the increase of $a_{h}$, $|\Delta z_h|$, $a_p$, and $P_{h,3}$ with $\bar{Z}$ in Table \ref{ta:structure}.

When dealing with two-dimensional materials, it has become customary to list polarization in units of C/m (see, {\em e.g.}, Refs.~\onlinecite{Hu2016,duerloo,fei_apl_2015_ges_gese_sns_snse,Mehboudi-prl}, among many others) or in Debye per unit cell (D/u.c.) as in Ref.~\onlinecite{ref1}. C/m units are employed here for direct comparison with the relevant literature.

\begin{table}[tb]
\caption {Optimized lattice constant $a_h$, signed buckling height $\Delta z_h$, and out-of-plane intrinsic polarization $P_{h,3}$ for group-IV monochalcogenide monolayers with a buckled honeycomb structure. The optimized lattice constant for a planar structure ($a_{p}$) is listed as well. Additional data from the literature was added for a direct comparison.}\label{ta:structure}
\begin{tabular}{c|c|ccc|cc}
\hline
Compound   & \scalebox{.7}{$\bar{Z}$} &$a_h$      & $\Delta z_h$ & $a_{p}$  & $P_{h,3}$ & $P_{h,3}$\\
           &                          &(\AA)      & (\AA)        & (\AA)    & (pC/m)    & (D/u.c.)\\
\hline
SiS  &15  &3.342        &$-$1.328      &4.242          & $-$0.7  & $-$0.02\\
SiSe &24  &3.551        &$-$1.425      &4.451          & $-$1.2  & $-$0.04\\
SiTe &33  &3.858        &$-$1.543      &4.758          & $-$2.0  & $-$0.08\\
GeS  &24  &3.524        &$-$1.370      &4.424          & 9.5    &    0.31\\
     &    &3.495$^*$    &$-$1.363$^*$  & ---           & ---     & ---    \\
     &    &3.489$^{\dagger}$ & ---     & ---           & ---     &    0.21$^{\dagger}$\\
     &    &3.495$^{\S}$ & ---          & ---           & ---     & ---    \\
GeSe &33  &3.697        &$-$1.466      &4.597          &  7.5    &    0.27\\
     &    &3.674$^*$    &$-$1.451$^*$  & ---           & ---     & ---    \\
     &    &3.663$^{\dagger}$ & ---     & ---           & ---     &    0.16$^{\dagger}$\\
     &    &3.676$^{\S}$ & ---          & ---           & ---     & ---    \\
GeTe &42  &3.980        &$-$1.579      &4.915          &  4.6    &    0.19\\
     &    &3.96$^{\ddag}$&$-$1.57$^{\ddag}$ & ---      & ---     & ---    \\
SnS  &33  &3.769        &$-$1.473      &4.719          & 13.3    &    0.49\\
     &    &3.753$^*$    &$-$1.465$^*$  & ---           & ---     & ---    \\
SnSe &42  &3.922        &$-$1.575      &4.922          & 9.3    &    0.37\\
     &    &3.910$^*$    &$-$1.564      & ---           & ---     & ---    \\
SnTe &51  &4.193        &$-$1.710      &5.243          &  7.3    &    0.33\\
PbS  &49  &3.944        &$-$1.468      &4.844          & 24.6    &    1.00\\
PbSe &58  &4.064        &$-$1.593      &5.014          & 21.5    &    0.93\\
PbTe &67  &4.323        &$-$1.722      &5.373          & 17.2    &    0.83\\
\hline
\end{tabular}\\
$^*$: PBE, {\em VASP}; Ref.~\onlinecite{Hu2016}. $^{\dagger}$: PBE, {\em VASP} (DF3 van der Waals corrections for few-layer stacks); Ref.~\onlinecite{ref1}.
 $^{\ddag}$: PBE, {\em VASP} (DF3 van der Waals corrections for few-layer stacks); Ref.~\onlinecite{GeTe}.
 $^{\S}$:    PBE, {\em VASP}; Ref.~\onlinecite{Gu2019}.
\end{table}

\section{Structural metastability}\label{stability}

\begin{figure*}[tb]
\includegraphics[width=0.96\textwidth]{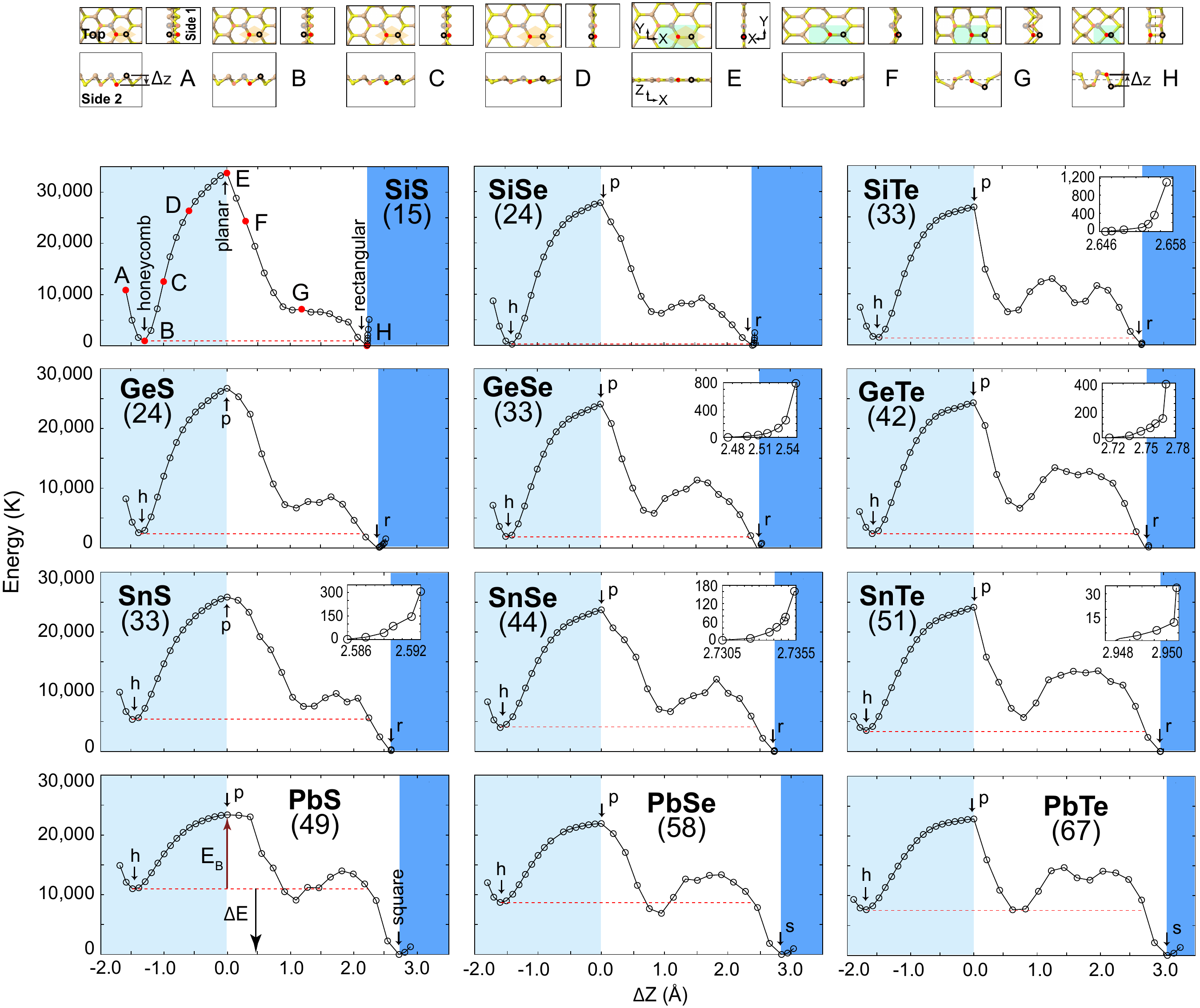}
\caption{Top: Selected atomistic configurations for a coordination-preserving transformation from a buckled honeycomb structure $A$ onto a unit cell with Pnm2$_1$ symmetry, as exemplified on a SiS monolayer. A rectangular cell containing four atoms is drawn in light green. Bottom: Energy {\em versus} $\Delta z$ for twelve two-dimensional group-IV monochalcogenide monolayers. $\Delta E$, $E_B$, and the locations along the path of the local minimum buckled honeycomb (h, structure B), planar (p, structure E), and ground state rectangular (r, structure H) unit cell with Pnm2$_1$ symmetry are highlighted. Insets show the energy cost to turn the rectangular unit cell into a square unit cell with Pmmn symmetry. PbS, PbSe, and PbTe have a ground state structure with Pmmn symmetry hosting a square (s) unit cell.}
\label{fig:fn4}
\end{figure*}

The methods illustrated in Fig.~\ref{fig:figure1} plus AIMD will now be applied to establish the structural metastability of the buckled phase of two-dimensional group-IV monochalcogenides.
\subsection{Local minimum and Kramers escape times}\label{stability1}

In direct analogy to the process we followed to determine the energy barrier $E_B$ underpinning the transformation from diamond to graphite, we first envision a structural transformation that preserves the three-fold atomistic coordination, and converts a structure with P3m1 symmetry onto the lower-energy Pnm2$_1$ one (or Pmmn for PbS, PbSe, and PbTe for reasons to be explained later on). This transformation is, in fact, the cis-to-trans conformal change alluded to in Fig.~\ref{fig:F1N}(a).

As it turns out, the transformation traverses through a planar structure ($\Delta z=0$) with p3m1 symmetry, whose lattice constant $a_p$ is listed in Table \ref{ta:structure} as well. The lack of inversion symmetry with respect to the $XY-$plane leads to a net intrinsic electric polarization $P_3$ for $\Delta z\leq 0$, whose magnitude is depicted in Fig.~\ref{fig:fn5}(b) ($X$, $Y$, and $Z$ axes are illustrated in Fig.~\ref{fig:fn4}, structure $E$). The in-plane three-fold and mirror symmetries of the P3m1 symmetry group are preserved in this structural transformation, rendering $P_1$ and $P_2$ equal to zero. (When piezolelectric properties are considered later on, the in-plane strain will break the in-plane three-fold symmetry, hence inducing a non-zero $P_1$ and $P_2$.)

The low-energy Pnm2$_1$ phase has the following symmetries: (i) the identity $E$; (ii) $\bar{C}_{2X}$: a two-fold rotation around the $X-$axis ($C_{2X}$), followed by a translation $\mathbf{R}=(\mathbf{a}'_1+\mathbf{a}'_2)/2$ where $\mathbf{a}'_1=(a_1',0,0)$ and $\mathbf{a}'_2=(0,a_2',0)$ are the lattice vectors of the rectangular cell (the magnitudes of $a_1'$ and $a_2'$ are provided as Supplemental Material in Ref.~\onlinecite{Shiva}); (iii) a glide-reflection plane $\bar{M}_{XY}$: a reflection by the $XY$ plane followed by $\mathbf{R}$; and (iv) a reflection about the $xz$ plane ($M_{XZ}$) \cite{rodin_prb_2016_sns}. The glide-reflection symmetry quenches the out-of-plane intrinsic polarization ($P_3=0$) for $\Delta z>0$, while the mirror symmetry along the $XZ$ plane renders $P_2=0$ for all values of $\Delta z$ in Fig.~\ref{fig:fn5}. Structures F to H show the process in which nearby atoms buckle in opposite directions (this is why the two-atom honeycomb structure turns into a four-atom unit cell), explaining in graphical terms why $P_3$ for $\Delta z \geq 0$. We used the pair of atoms highlighted in solid red and open black colors to define $\Delta z$; this pair swaps the sign $\Delta z$ on structures F to H with respect to its value in structures A to D, hence giving the change of sign on the $\Delta z$ parameter we used to describe the transformation.

A transformation of the Pnm2$_1$ unit cell into a four-fold symmetric (square, s) Pmmn structure with a net $\mathbf{P}=0$ takes place within the dark-blue section of the energy {\em versus} $\Delta z$ plots for nine compounds (SiS, SiSe, SiTe, GeS, GeSe, GeTe, SnS, SnSe, and SnTe)\cite{Mehboudi-nl,Salvador2018-prb,Shiva,Villanova2020PRB} for which the Pmmn structure has a larger energy, as verified by the increase in energy in Fig.~\ref{fig:fn4}. Zoom-ins were inserted to some subplots to emphasize the increase in energy. Two-dimensional PbS, PbSe, and PbTe display a global minimum with Pmmn symmetry already;\cite{Mehboudi-nl,Salvador2018-prb,Shiva,Villanova2020PRB} and $\Delta z$ was just enlarged in these structures to highlight such (global) minima. More specific technical details of the transformation can be found as Supplemental Material\cite{suppl}. Here, it suffices to say that the energies shown in Fig.~\ref{fig:fn4} take on their minimum possible magnitudes for fixed values of $\Delta z$, following an explicit optimization procedure similar to the one employed in Fig.~\ref{fig:figure1}.

Summarizing, the energy {\em versus} $\Delta z$ plots are divided in three sections. The first one, colored in light blue, corresponds to the P3m1 to p3m1 (buckled honeycomb to planar honeycomb) part of the transformation. The second portion, in white color, takes the p3m1 structure onto either the Pnm2$_1$ structure (for SiS, SiSe, SiTe, GeS, GeSe, GeTe, SnS, SnSe, and SnTe) or the Pmmn structure (for PbS, PbSe, and PbTe). Figure \ref{fig:fn4} demonstrates, without exception, that the energy barrier $E_B$ takes its maximum value for the planar structure. Values of $E_B$, $\Delta E$, and $\omega$ are listed in Table \ref{ta:ta1}. We note that $E_B$ is of the same order of magnitude than the one listed for the diamond to graphite transformation, and Eqn.~\eqref{eq:escape} yields escape times ranging from a month for PbS, up to multiple ages of the universe for SnTe, GeTe, GeSe, SiTe, SiSe, and SiS monolayers with a buckled honeycomb structure (Table \ref{ta:ta2}).

\begin{table}[tb]
\caption{Energy difference $\Delta E$, energy barrier $E_B$, and $\omega$ for two-dimensional group-IV monochalcogenide compounds.}\label{ta:ta1}
\begin{tabular}{c|c|cc|c}
\hline
Compound              &\scalebox{.7}{$\bar{Z}$}            &  $\Delta E$ (K) & $E_B$ (K) & $\omega$ (THz) \\
\hline
P3m1 to Pnm2$_1$ SiS  &15     & $-$858          & 32,927 & 9.1\\
P3m1 to Pnm2$_1$ SiSe &24     & $-$115          & 27,749 & 6.6\\
P3m1 to Pnm2$_1$ SiTe &33     & $-$1,271        & 25,630 & 5.1\\
P3m1 to Pnm2$_1$ GeS  &24     & $-$2,511        & 24,230 & 6.4\\
P3m1 to Pnm2$_1$ GeSe &33     & $-$1,794        & 22,282 & 5.2\\
P3m1 to Pnm2$_1$ GeTe &42     & $-$2,344        & 21,849 & 4.2\\
P3m1 to Pnm2$_1$ SnS  &33     & $-$5,358        & 20,461 & 4.7\\
P3m1 to Pnm2$_1$ SnSe &42     & $-$3,979        & 19,731 & 4.1\\
P3m1 to Pnm2$_1$ SnTe &51     & $-$3,512        & 20,617 & 3.5\\
P3m1 to Pmmn PbS      &49     & $-$11,016       & 12,393 & 3.4\\
P3m1 to Pmmn PbSe     &58     & $-$9,023        & 13,194 & 3.0\\
P3m1 to Pmmn PbTe     &67     & $-$7,625        & 15,185 & 2.7\\
\hline
\end{tabular}
\end{table}

\begin{table}[tb]
\caption{Estimated escape times $\tau$ for group-IV monochalcogenides with a (P3m1) buckled honeycomb structure. (The age of the universe is of the order of $10^{10}$ years.)}\label{ta:ta2}
\begin{tabular}{c|c|c||c|c|c}
\hline
Compound &\scalebox{.7}{$\bar{Z}$} & $\tau$ (years) & Compound &\scalebox{.7}{$\bar{Z}$} & $\tau$ (years)\\
\hline
SiS         &15              & 1.0$\times 10^{28}$              &SnS         &33              & 1.8$\times 10^{10}$  \\
SiSe        &24              & 4.5$\times 10^{20}$              &SnSe        &42              & 1.8$\times 10^{9}$   \\
SiTe        &33              & 5.0$\times 10^{17}$              &SnTe        &51              & 4.0$\times 10^{10}$  \\
GeS         &24              & 3.7$\times 10^{15}$              &PbS         &49              & 0.05                  \\
GeSe        &33              & 7.0$\times 10^{12}$              &PbSe        &58              & 0.80                  \\
GeTe        &42              & 2.0$\times 10^{12}$              &PbTe        &67              & 7.1$\times 10^{2}$   \\
\hline
\end{tabular}
\end{table}

\subsubsection{Evolution of polarization along the structural transformation}

\begin{figure}[tb]
\includegraphics[width=0.48\textwidth]{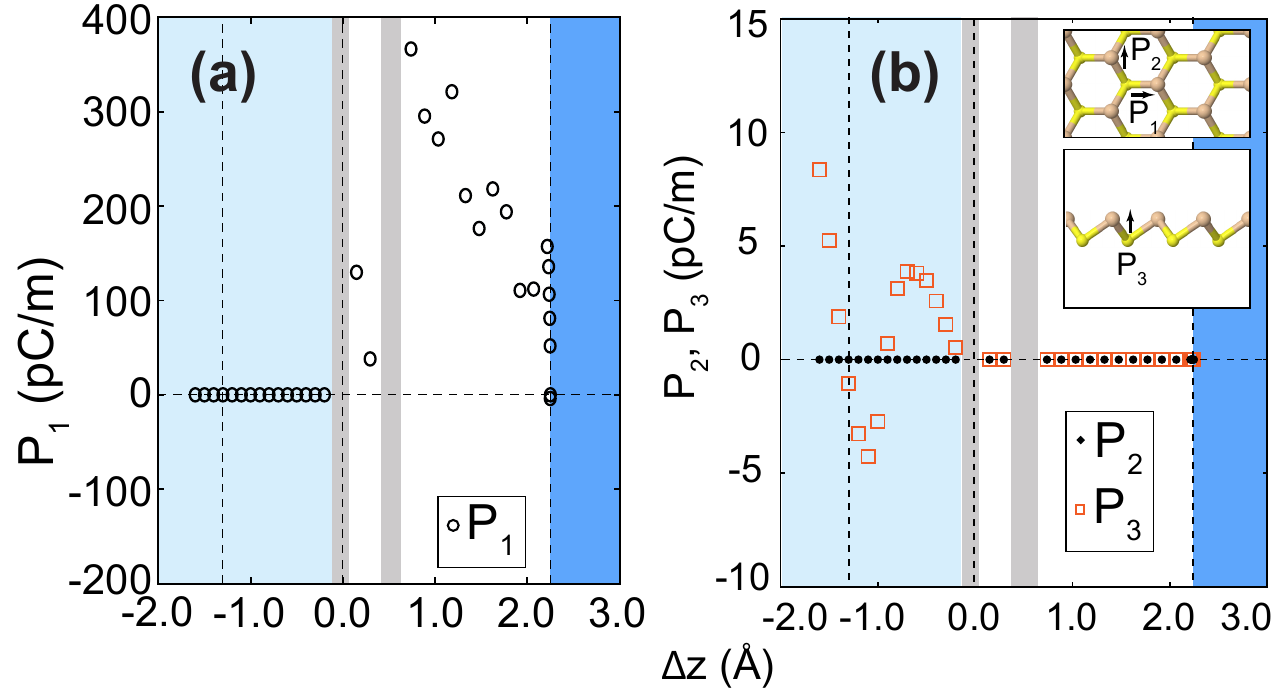}
\caption{Evolution of the intrinsic electric polarization $\mathbf{P}=(P_1,P_2,P_3)$ for a SiS monolayer along the structural transformation depicted in Fig.~\ref{fig:fn4}. The light and bold blue shading are consistent with the structural evolution depicted in Fig.~\ref{fig:fn4}. From left to right, the vertical dashed lines correspond to the local minima buckled honeycomb structure (h), the planar structure with p3m1 symmetry (p), and the rectangular structure with Pnm2$_1$ symmetry (r). The directions of the polarization vector are displayed as an inset in subplot (b).}
\label{fig:fn5}
\end{figure}

Figure \ref{fig:fn5} depicts the evolution of the intrinsic electric polarization $\mathbf{P}=(P_1,P_2,P_3)$ of the SiS monolayer along the $\Delta z$ path depicted in Fig.~\ref{fig:fn4}. From left to right, vertical dashed lines label structures h, p, and r (B, E, and H in Fig.~\ref{fig:fn4}, respectively). Structures with a honeycomb unit cell ($\Delta z\leq 0$) are three-fold symmetric and hence lack an intrinsic in-plane electric polarization ($P_1=P_2=0$ in Fig.~\ref{fig:fn5}).\cite{C3CP53971G} Shaded vertical areas indicate structures that yield a metallic electronic structure\cite{suppl} for which the polarization cannot be computed within the standard approach,\cite{modern} leading to a slight discontinuity for $P_1$ for $\Delta z$ in between 0 and 0.5 \AA{}.

Nevertheless, and as indicated in Ref.~\onlinecite{ref1} and observed in Fig.~\ref{fig:fn5}(b), an out-of-plane intrinsic polarization $P_3$ ensues in binary compounds with P3m1 group symmetry. The magnitude of $P_3$ at the local buckled honeycomb energy minimum is labeled $P_{h,3}$ and its magnitude is listed in Table \ref{ta:structure} for the twelve studied compounds. In order to understand the sign of $P_3$, we remark that all compounds with a P3m1 symmetry have their chalcogen atoms below group-IVA atoms. As seen in Fig.~\ref{fig:fn5}(b), the magnitude of $P_3$ on the P3m1 phase is tunable by a change in structure ($\Delta z$). This fact is an incipient demonstration of out-of-plane piezoelectric behavior, a discussion omitted in previous work\cite{Hu2016} that will be provided in Sec.~\ref{piezo}. Belonging to another symmetry group, structures labeled with $\Delta z>0$ suppress their out-of-plane polarization $P_3$, and may develop a net in-plane polarization $P_1$.\cite{Mehboudi-prl} [The exception is the Pmmn (s) phase for which
$P_1=P_2=P_3=0$.\cite{Mehboudi-prl,Salvador2018-prb,Shiva,Villanova2020PRB}]

\begin{figure*}[tb]
\includegraphics[width=0.96\textwidth]{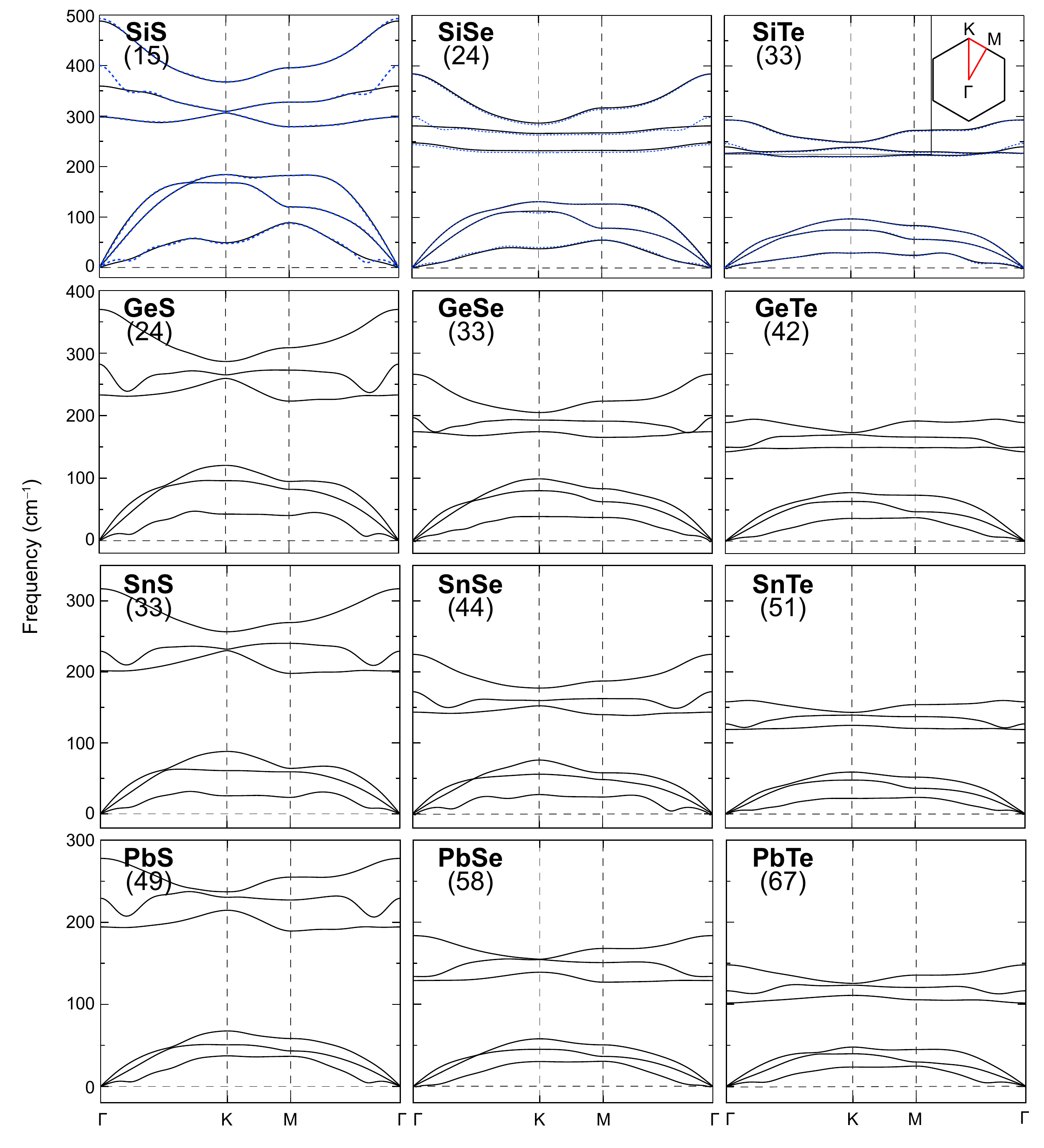}
\caption{Phonon dispersion for the twelve group-IV monochalcogenide monolayers with a buckled honeycomb structure at the local energy minima (h) depicted in Fig.~\ref{fig:fn4}: no negative modes can be seen, and these results constitute the second proof of the structural stability of group-IV monochalcogenide monolayers with a buckled honeycomb structure. The blue dashed curves were obtained with dipole corrections.}
\label{fig:fn6}
\end{figure*}

\begin{figure*}[tb]
\includegraphics[width=0.8\textwidth]{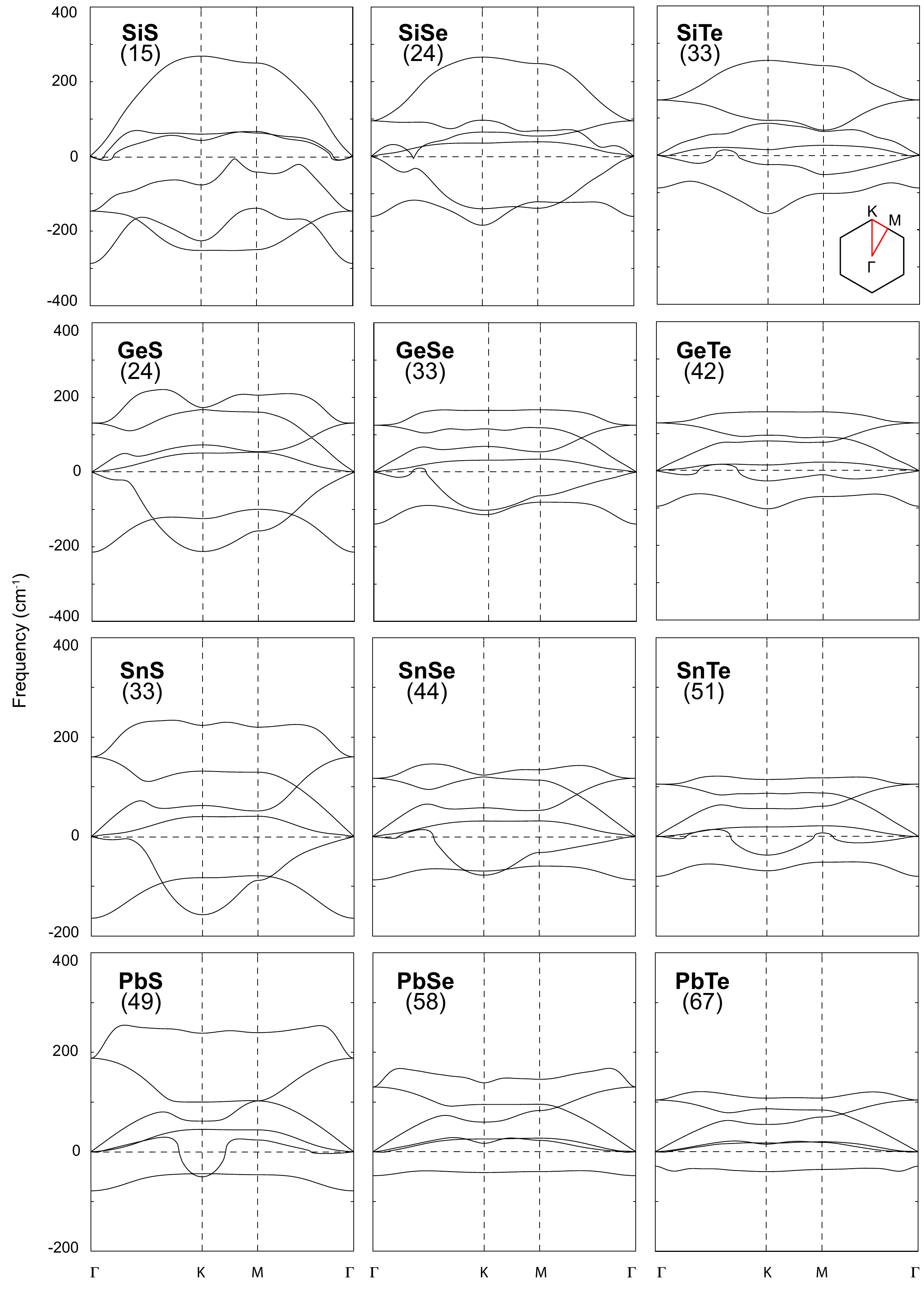}
\caption{Phonon dispersion for the twelve group-IV monochalcogenide monolayers with a planar honeycomb structure at the energy maxima (p) depicted in Fig.~\ref{fig:fn4} (no dipole corrections are required given that the structure is planar): With the exception of SiS, all materials show a single negative mode at the $\Gamma$-point, corresponding with an out-of-plane optical vibration, thus verifying that point $p$ indeed corresponds with the maximum value of the energy barrier.}
\label{fig:fn6bis}
\end{figure*}

\subsection{Phonon dispersion calculations}\label{stability2}
The phonon dispersions shown in Fig.~\ref{fig:fn6} were computed at the local energy minima (labeled h) in Fig.~\ref{fig:fn4}. {The lower band at the $\Gamma-$point does not have a quadratic dispersion, like the one seen in graphene. There are two reasons for such a discrepancy. Direct inspection into (i) the force constant tensor and (ii) into the disposition of atoms in the supercell from which phonons are computed permits observing the force matrices for pairs of atoms lying along the $x-$axis explicitly---and corresponding to first, third, fifth, and eight nearest neighbors. These matrices are diagonal in graphene (see Equation 9.9 in Ref.~\onlinecite{Dresselhaus}), but they couple substantially along the $x-$ and $z-$directions in group-IV monochalcogenide monolayers with a buckled honeycomb structure. For instance, the matrix coupling among the central Si atom and a first-neighbor S atom (located to its left along the $x-$axis and a distance $|\Delta z_h|$ above) is given by:
\begin{equation*}
K_{SiS}(1)=\begin{pmatrix}
-2.7497	& 0.0000 & 3.6220\\
 0.0000	&-0.8307 & 0.0000\\
 1.8594	& 0.0000 &-3.6082\\
           \end{pmatrix}
\end{equation*}
in units of eV/\AA${}^{2}$. The coupling among in-plane and out-of-plane directions has an obvious origin: a horizontal relative displacement necessarily modifies the relative height among atoms, and viceversa. In graphene, on the other hand, a horizontal displacement does not necessarily require a change on the relative vertical distances among atoms.  The second reason for the discrepancy among the dispersion of the lowest energy phonon band and the one seen in graphene has to do with the low wavelength behavior, which couples to the long-range electrostatic interaction among ions in these binary materials:} as indicated in Sec.~\ref{sec:cm}, the effect of Born charges turned crucial to describe the long-wavelength vibrational behavior appropriately.\cite{PhysRevB.1.910} Lacking negative frequencies, and providing additional credence to the escape time argument provided in Table \ref{ta:ta2}, the phonon dispersion calculations in Fig.~\ref{fig:fn6} represent the second proof of structural metastability of group IV monochalcogenide monolayers with a buckled honeycomb structure.

{ It is illustrative to look at the vibrational spectra at the top of the barrier (point p in Fig.~\ref{fig:fn4}), and Fig.~\ref{fig:fn6bis} serves this purpose. All materials, with the exception of SiS, show a single negative mode at the $\Gamma$-point, corresponding with an out-of-plane optical vibration, and thus verifying that point $p$ indeed corresponds with the maximum value of the energy barrier. SiS was known to have another ground state since earlier versions of this manuscript (Reference \onlinecite{yang_nanolett_2015_sis}); that compound displays an out-of-plane optical vibration, and two degenerate in-plane vibrations accordingly.}

\subsection{{\em Ab initio} molecular dynamics}\label{stability3}
Table \ref{ta:ta1} indicates that SiS has the highest energy barrier $E_B$, and PbS has the lowest one among the twelve compounds studied. Therefore, if PbS turns to be metastable at room temperature in AIMD calculations, one may be able to conclude that the remaining eleven compounds should be metastable, too. Such assertion could be checked by performing AIMD on a second compound, say SiS.

It is important to be judicious when choosing the type of molecular dynamics calculations to be performed: structural transformations that require the materials' area to change---such as the one illustrated in Fig.~\ref{fig:fn4}---may display a forced stability in NVT AIMD calculations at a given temperature, because of the constant area constraint, that is inconsistent with the structural transformation (observe the change in area on the top views in Fig.~\ref{fig:fn4}). In the structural transformations studied by this team, we have shown that structural instabilities are triggered at lower temperatures when employing the NPT ensemble,\cite{Villanova2020PRB} as the area of these 2D materials is allowed to change. This explains our choice of an NPT ensemble for the AIMD calculations reported here.

\begin{figure}[tb]
\includegraphics[width=0.48\textwidth]{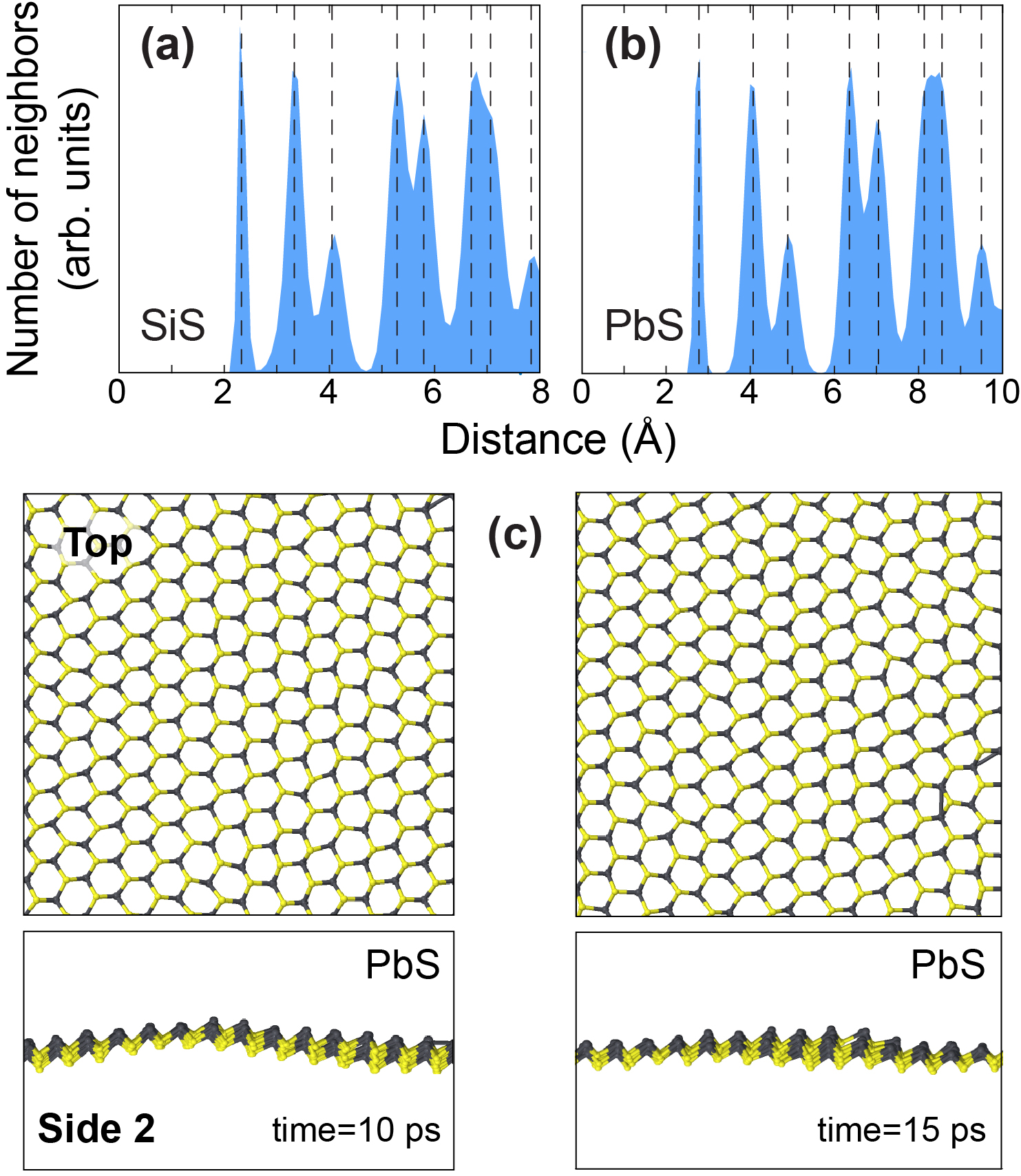}
\caption{Distribution of interatomic distances at 300 K for the (a) SiS monolayer and (b) PbS monolayer with a buckled honeycomb structure. The dashed vertical lines---which match the distribution's peaks---are interatomic distances at zero temperature. AIMD thus confirms the structural metastability of these buckled honeycomb phases at room temperature.}
\label{fig:fn7}
\end{figure}

\begin{table}[tb]
\caption{Melting temperature $T_M$ of bulk group-IV monochalcogenides according to Ref.~\onlinecite{Landolt}. (Data for Si-based compounds is lacking.)}\label{ta:melt}
\begin{tabular}{c|c|c||c|c|c}
\hline
\hline
Compound & \scalebox{.7}{$\bar{Z}$}& $T_M$ (K) & Compound & \scalebox{.7}{$\bar{Z}$}& $T_M$ (K)\\
\hline
\hline
GeS  &24    & 938     & SnTe &51    & 1,063 \\
GeSe &33    & 948     & PbS  &49    & 1,383 \\
GeTe &42    & 997     & PbSe &58    & 1,355 \\
SnS  &33    & 1,153   & PbTe &67    & 1,197 \\
SnSe &42    & 1,153    \\
\hline
\end{tabular}
\end{table}

The distributions of interatomic distances at 300 K for NPT AIMD calculations on SiS or PbS supercells containing 336 atoms, and running for up to 15 ps, are shown in Fig.~\ref{fig:fn7}. The peaks of these distributions match the zero-temperature interatomic distances shown by vertical dashed lines. The supercell size and runtimes in our calculations are much larger than these employed in earlier works,\cite{Hu2016,ref1} and these AIMD results contribute the third and final demonstration of structural metastability of this family of compounds at room temperature.

\begin{figure}[tb]
\includegraphics[width=0.48\textwidth]{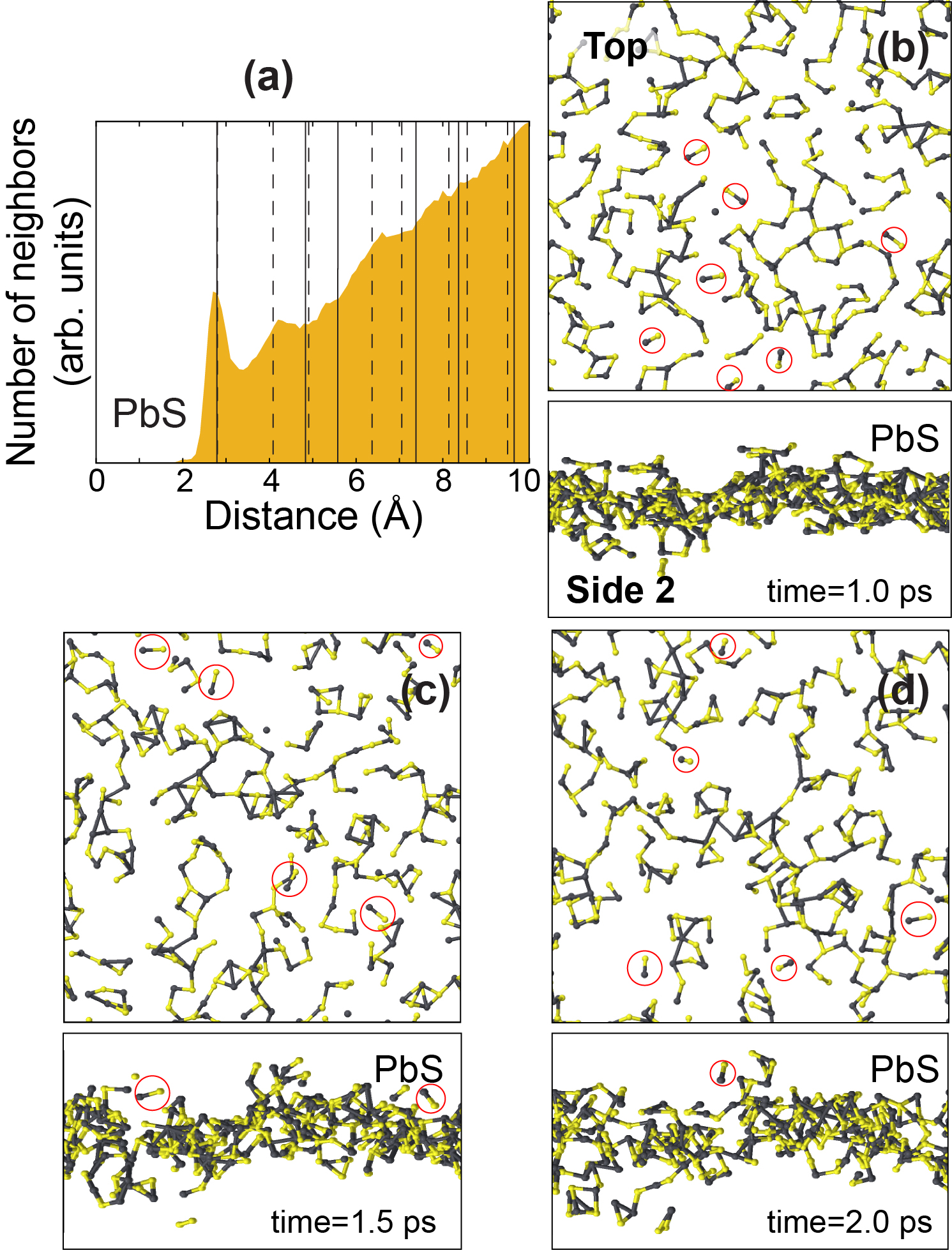}
\caption{(a) Distribution of interatomic distances at 2,000 K for the PbS monolayer with a buckled honeycomb structure. The dashed vertical lines are interatomic distances at zero temperature, and the solid vertical lines correspond to interatomic distances of the planar structure. The sharp peak at 2.8 \AA{} is a sign of dimerization, that is confirmed by the presence of dimers within red circles in structural snapshots at 1.0 (subplot b), 1.5 (subplot c), and 2.0 ps (subplot d), respectively: PbS turns amorphous at T=2,000 K.}
\label{fig:8}
\end{figure}

\begin{figure*}[tb]
\includegraphics[width=0.96\textwidth]{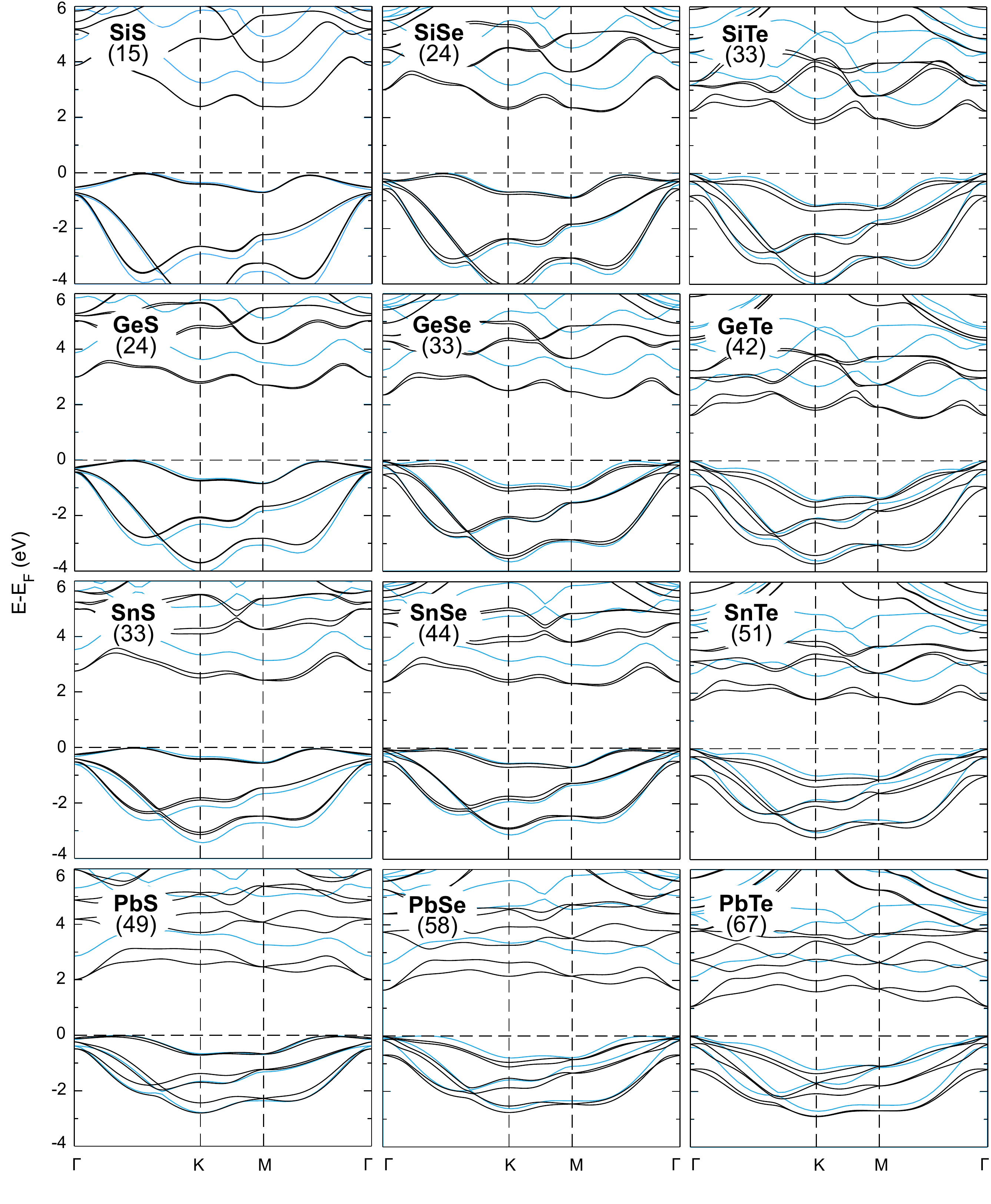}
\caption{Electronic band structures with spin-orbit coupling included (in black) for the twelve group-IV monochalcogenide monolayers with a buckled honeycomb structure at the local energy minima (h) depicted in Fig.~\ref{fig:fn4}. The band structures in cyan were obtained with the HSE06 functional and without spin-orbit coupling.}
\label{fig:fn9}
\end{figure*}

The next question concerns the feasibility of realizing the elastic structural transformation depicted in Fig.~\ref{fig:fn4} thermally: after all, the Pnm2$_1$ to Pmmn transformation can be achieved with temperature.\cite{Mehboudi-nl,Chang2016,Mehboudi-prl,Salvador2018-prb,Villanova2020PRB} Figure \ref{fig:8} shows that the barrier is too high, such that thermal fluctuations in an AIMD calculation at 2,000 K turn the honeycomb PbS structure onto an amorphous material that even undergoes dimerization. This result appears to be consistent with a melting temperature $T_M$ not larger than 1,200 K (Table \ref{ta:melt}) for the corresponding bulk phase. One then concludes that the energy barrier $E_B$ is too tall for all twelve compounds for the elastic transformation depicted in Fig.~\ref{fig:fn4} to take place solely by temperature.

\section{Electronic band structures}\label{bands}

Now that the metastability of these honeycomb phases has been demonstrated using three independent methods, it is time to turn to the electronic properties of these compounds, which were obtained with the spin-orbit coupling turned on (in black), and with the HSE06 hybrid functional \cite{HSE06} without spin-orbit coupling turned on (in cyan) and displayed in Fig.~\ref{fig:fn9}. (HSE06 corrections provide bandgaps in closer agreement with experimental estimates.) With the exception of PbSe and PbTe, these P3m1 two-dimensional phases feature indirect bandgaps, and DFT band gaps within 2.12 and 3.40 eV when HSE06 corrections are considered (see Table \ref{ta:taGaps}): the electronic band gap is a relevant parameter for applications argued for in previous work,\cite{ref1,Gu2019} and its magnitude can be further tuned by the choice of chemical elements in these compounds.

\begin{table}[tb]
\caption{Electronic band gaps $E_g$ and $E_g^{HSE}$ for group-IV monochalcogenides with a (P3m1) buckled honeycomb structure. When available, additional data from the literature was added for a direct comparison.}\label{ta:taGaps}
\begin{tabular}{c|c|c|c}
\hline
Compound &\scalebox{.7}{$\bar{Z}$} & $E_g$ (eV) & $E_g^{HSE}$ (eV)\\
\hline
SiS         &15              & 2.41             & 3.24\\
SiSe        &24              & 2.23             & 3.03\\
SiTe        &33              & 1.65             & 2.47\\
GeS         &24              & 2.59, 2.47$^*$, 2.49$^{\S}$& 3.40, 3.27$^{\dagger}$, 3.27$^{\S}$ \\
GeSe        &33              & 2.27, 2.27$^*$, 2.29$^{\S}$& 3.08, 3.01$^{\dagger}$, 2.99$^{\S}$ \\
GeTe        &42              & 1.56,            & 2.34, 2.35 $^{\ddag}$  \\
SnS         &33              & 2.43, 2.31$^*$   & 3.14\\
SnSe        &42              & 2.27, 2.21$^*$   & 2.97\\
SnTe        &51              & 1.62             & 2.44\\
PbS         &49              & 2.03             & 2.85\\
PbSe        &58              & 1.68             & 2.59\\
PbTe        &67              & 1.10             & 2.12\\
\hline
\end{tabular}\\
$^*$: PBE, {\em VASP}; Ref.~\onlinecite{Hu2016}. $^{\dagger}$: PBE, HSE06 functional for electronic properties, {\em VASP} (DF3 van der Waals corrections for few-layer stacks); Ref.~\onlinecite{ref1}. $^{\ddag}$: PBE, HSE06 functional for electronic properties, {\em VASP} (DF3 van der Waals corrections for few-layer stacks); Ref.~\onlinecite{GeTe}.
 $^{\S}$:  both  PBE and HSE06 results were reported using {\em VASP}; Ref.~\onlinecite{Gu2019}.
\end{table}

\section{Elastic and piezoelectric properties}\label{piezo}

The in-plane symmetry operations of hexagonal boron nitride ({\em wallpaper}, two-dimensional, symmetry group p3m1) apply to the buckled hexagonal compounds, but the latter lacks out-of-plane inversion symmetry and thus belong to the ({\em three-dimensional}) symmetry group P3m1 (Sec.~\ref{structure}). The three-fold symmetry of these compounds renders $C_{22}=C_{11}$,\cite{duerloo} and $C_{66}=\frac{C_{11}-C_{12}}{2}$.\cite{c66} ($C_{66}=C_{1212}$ is the elastic coefficient due to in-plane shear strain $\epsilon_{12}$, $C_{21}=C_{12}$, and Voigt notation has been employed throughout.) The magnitudes of an elastic coefficients  $C_{11}$ and $C_{12}$ are calculated using a fully relaxed atomic configuration( relaxed ion) and listed in Table \ref{ta:ta4}.

\begin{figure*}[tb]
\includegraphics[width=0.9\textwidth]{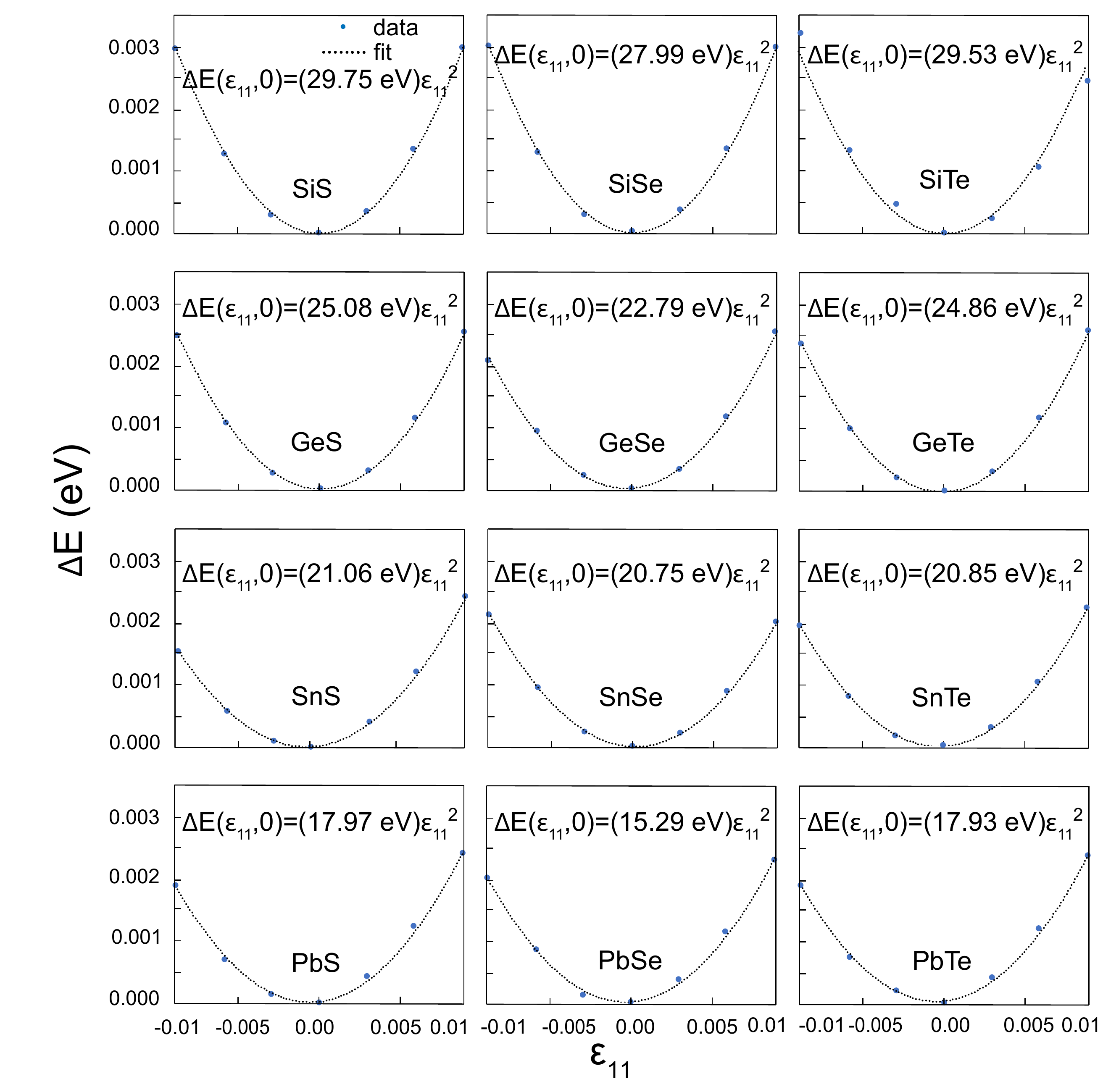}
\caption{Quadratic fitting of energy versus uniaxial strain $\epsilon_{11}$. The coefficient shown at each plot is $2C_{11}$.}\label{fig:FN10}
\end{figure*}

Elastic constants are calculated following the procedure established by Duerloo and coworkers:\cite{duerloo} A rectangular unit cell with four atoms is set, and the following expression is employed:
\begin{equation}
\Delta E(\epsilon_{11},\epsilon_{22})=\frac{C_{11}}{2}\left(\epsilon_{11}^2+\epsilon_{22}^2\right)+C_{12}\epsilon_{11}\epsilon_{22}.
\end{equation}
$C_{11}$ is obtained by applying uniaxial strain along the $\epsilon_{11}$ direction, which yields:
\begin{equation}\label{eq:eq3}
\Delta E(\epsilon_{11},0)=\frac{C_{11}}{2}\epsilon_{11}^2.
\end{equation}
The quadratic coefficient in Eq.~\eqref{eq:eq3} shown in Fig.~\ref{fig:FN10} is thus $C_{11}/2$.

\begin{figure*}[tb]
\includegraphics[width=0.9\textwidth]{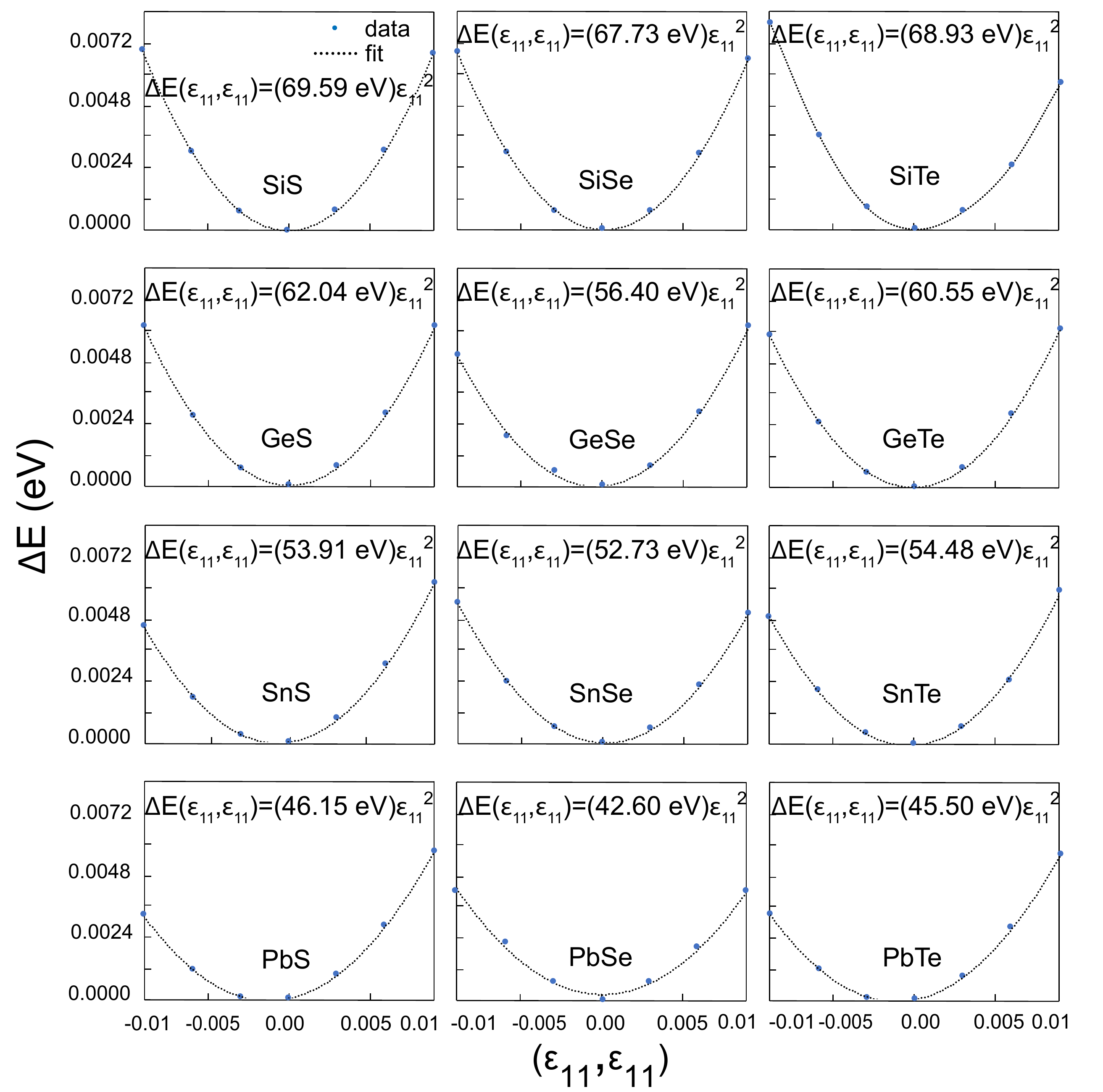}
\caption{Quadratic fitting of energy versus biaxial strain.}
\label{fig:FN11}
\end{figure*}

$C_{12}$, in turn, is obtained by applying biaxial strain ($\epsilon_{11},\epsilon_{22}$) with $\epsilon_{22}=\epsilon_{11}$, which yields:
\begin{equation}\label{eq:eq4}
\Delta E(\epsilon_{11},\epsilon_{11})=C_{11}\epsilon_{11}^2+C_{12}\epsilon_{11}^2=2\Delta E(\epsilon_{11},0)+C_{12}\epsilon_{11}^2,
\end{equation}
where the presence of a contribution from uniaxial strain is made explicit. A quadratic fitting of Eq.~\eqref{eq:eq4} yields $\Delta E(\epsilon_{11},\epsilon_{11})=\alpha\epsilon_{11}^2$ as shown in Fig.~\ref{fig:FN11}. Using Eqns.~\eqref{eq:eq3} and \eqref{eq:eq4} we get:
\begin{equation}
C_{12}\epsilon_{11}^2=\Delta E(\epsilon_{11},\epsilon_{11})-2\Delta E(\epsilon_{11},0)=(\alpha-C_{11})\epsilon_{11}^2.
\end{equation}

It is customary practice to divide these coefficients by the equilibrium area of the unit cell, and to write these constants in units of N/m. The area of the four atom unit cell at the local minimum is given by $\sqrt{3}a_{h}^2$, with $a_h$ given in Table \ref{ta:structure} . The constants obtained from the quadratic fitting as shown Fig. \ref{fig:FN10} and Fig. \ref{fig:FN11} are thus multiplied by $\frac{1.602 \times 10^{-19} C/e}{\sqrt{3}a_h^2}$ to get the values listed in Table \ref{ta:ta4}.

Elastic and piezoelectric properties have a strong dependence on the choice of exchange-correlation functional, because forces and dipole moments depend directly on the charge distribution. Even though it is the default choice for many calculations, the PBE functional was never developed to account for materials with a large amount of vacuum, while {\em ab initio} vdW functionals are explicitly developed to account for drastic changes in density at interstitials and vacuum regions.\cite{reviewvdw} As indicated before, the elastic coefficients reported in Table \ref{ta:ta4} were obtained under {\em relaxed-ion} conditions; {\em i.e.,} atomic positions were optimized as strain was applied.

\begin{table}[tb]
\caption{{Relaxed-ion}, in-plane elastic coefficients $C_{11}$ and $C_{12}$ (in N/m) for group-IV monochalcogenides with a (P3m1) buckled honeycomb structure.  When available, additional data from the literature was added for a direct comparison.}\label{ta:ta4}
\begin{tabular}{c|c|cccc}
\hline
Compound & \scalebox{0.7}{$\bar{Z}$} && $C_{11}$ && $C_{12}$ \\
\hline
SiS      & 15     && 49.37                && 8.36   \\
SiSe     & 24     && 40.88                && 8.59   \\
SiTe     & 33     && 36.64                && 6.62   \\
GeS      & 24     && 37.34, 50.63$^*$     && 8.85, 10.78$^*$ \\
GeSe     & 33     && 29.73, 49.54$^*$     && 7.29, 10.42$^*$ \\
GeTe     & 42     && 29.04     && 6.33   \\
SnS      & 33     && 27.39, 41.59$^*$     && 7.66, 10.52$^*$ \\
SnSe     & 42     && 24.27, 38.99$^*$     && 6.74,  8.56$^*$ \\
SnTe     & 51     && 21.77     && 6.72 \\
PbS      & 49     && 21.33     && 6.05 \\
PbSe     & 58     && 17.04     && 5.28 \\
PbTe     & 67     && 17.77     && 4.77 \\
\hline
\end{tabular}\\
$^*$: PBE, {\em VASP}; Ref.~\onlinecite{Hu2016}.
\end{table}

\begin{figure}[tb]
\includegraphics[width=0.48\textwidth]{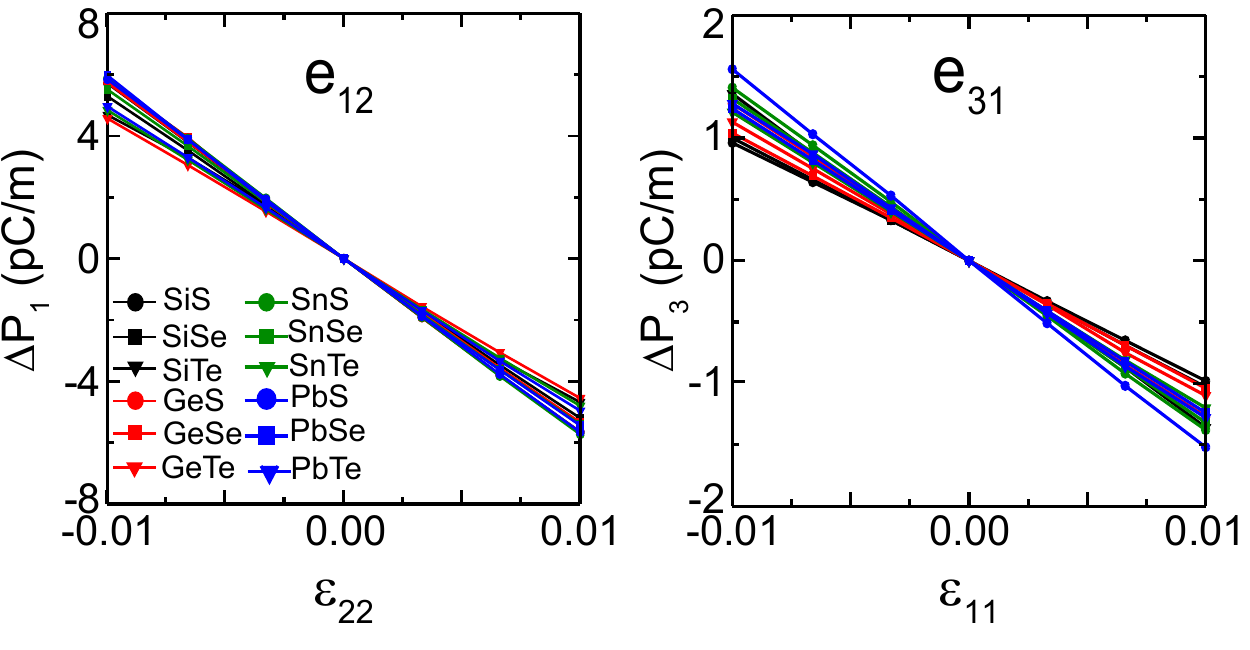}
\caption{Non-zero piezoelectric responses of group-IV monochalcogenide monolayers with a buckled honeycomb structure.}
\label{fig:fn12}
\end{figure}

Materials with a P3m1 group symmetry follow the in-plane symmetry rules that were developed in Ref.~\onlinecite{duerloo}, but out-of-plane piezoelectric properties ought to be added to the discussion of these novel phases. The symmetries of the P3m1 group render the piezoelectric coefficients $e_{21}$,  $e_{22}$, $e_{16}$, and $e_{36}$ equal to zero because mirror symmetry is preserved upon the application of normal strain, and because the in-plane and out-of-plane polarization are not sensitive to the sign (direction) of the in-plane shear strain. The $e_{11}$ piezoelectric coefficient is different from zero, and $e_{12}=e_{26}=-e_{11}$. Here, the atomistic structure is defined so that $e_{11}$ is positive, as in Ref.~\onlinecite{duerloo} and $e_{12}$ is shown in Fig.~\ref{fig:fn9} for a direct comparison with the PBE results from Ref.~\onlinecite{Hu2016} (in which an overall negative sign appears related to their choice of $X$ direction, which corresponds to the $-X$ direction in Ref.~\onlinecite{duerloo} and in the present work). The remaining non-zero coefficients are the out-of-plane piezoelectric responses $e_{31}=e_{32}$, which were not studied in previous work.\cite{Hu2016}

The $e_{31}$ coefficient is negative because the buckling height $\Delta z$ turns closer to zero for tensile strain, hence quenching the intrinsic electric polarization [see structures A to E in Fig.~\ref{fig:fn4}, and $P_3$ in Fig.~\ref{fig:fn5}(b)]. $\Delta P_i$ {\em versus} strain plots---whose slopes are the non-zero piezoelectric coefficients $e_{11}$ and $e_{31}$ listed in Table \ref{ta:piezo}---are shown in Fig.~\ref{fig:fn9}.

\begin{table}[]
\caption{Relaxed-ion piezoelectric coefficients  $e_{12}$ and $e_{31}$ for group-IV monochalcogenides with a (P3m1) buckled honeycomb structure. Direct piezoelectric coefficients $d_{12}$ and $d_{31}$ are listed as well.   When available, additional data from the literature was added for a direct comparison.}\label{ta:piezo}
\begin{tabular}{c|c|cc|cc}
\hline
Compound & \scalebox{0.7}{$\bar{Z}$} & $e_{12}$ ($\frac{\text{pC}}{\text{m}}$) &$e_{31}$ ($\frac{\text{pC}}{\text{m}}$) & $d_{12}$ ($\frac{\text{pm}}{\text{V}}$)  &$d_{31}$ ($\frac{\text{pm}}{\text{V}}$)\\
\hline
SiS    &15  &     $-$558     &$-$92        &  $-$13.60&$-$2.24 \\
SiSe   &24  &     $-$515     &$-$95       &  $-$15.95&$-$2.93 \\
SiTe   &33  &     $-$425     &$-$90       &  $-$14.17&$-$3.00 \\
GeS    &24  &     $-$560     &$-$121       &  $-$19.67&$-$4.26 \\
       &    &       225$^*$  & ---         &  5.65$^*$& ---    \\
GeSe   &33  &     $-$554     &$-$103       &  $-$24.67&$-$4.58 \\
       &    &       191$^*$  & ---         &  4.88    & ---    \\
GeTe   &42  &     $-$449     &$-$101       &  $-$19.78&$-$4.43 \\
SnS    &33  &     $-$574     &$-$135       &  $-$29.07&$-$6.82 \\
       &    &       164      & ---         &  5.28    & ---    \\
SnSe   &42  &     $-$537     &$-$125       &  $-$30.67&$-$7.12 \\
       &    &       141      & ---         &  4.63    & ---    \\
SnTe   &51  &     $-$475     &$-$116       &  $-$31.56&$-$7.68 \\
PbS    &49  &     $-$567     &$-$150       &  $-$37.15&$-$9.79 \\
PbSe   &58  &     $-$550     &$-$145       &  $-$46.72&$-$12.30 \\
PbTe   &67  &     $-$490     &$-$126       &  $-$37.72&$-$9.68 \\
\hline
\end{tabular}\\
$^*$: PBE, {\em VASP}; Ref.~\onlinecite{Hu2016}.
\end{table}

The non-zero direct piezoelectric coefficients are $d_{11}=\frac{e_{11}C_{11}-e_{12}C_{12}}{C_{11}^2-C_{12}^2}$, $d_{12}=\frac{e_{12}C_{11}-e_{11}C_{12}}{C_{11}^2-C_{12}^2}$, $d_{26}=-\frac{e_{11}}{C_{66}}$, and $d_{31}=\frac{e_{31}}{C_{11}+C_{12}}$ ($d_{32}=d_{31}$). The magnitudes of $d_{12}$ and $d_{31}$ are listed in Table \ref{ta:piezo}, too. Coefficients $e_{31}$ and $d_{31}$ supplement the in-plane coefficients reported in previous work.\cite{Hu2016}

\section{Conclusions}\label{conclusions}

The following contributions are contained in the present work: (a) We determined the group symmetry of group-IV monochalcogenide monolayers with a buckled honeycomb structure. (b) We expanded the list of metastable group-IV monochalcogenide monolayers with a honeycomb structure to all twelve compounds on this material family: SiS, SiSe, SiTe, GeS, GeSe, GeTe, SnS, SnSe, SnTe, PbS, PbSe, PbTe. (c) The electronic band structure and electronic band gap were contributed, too. (d) We also contributed the out-of-plane intrinsic electric polarization and the piezoelectric response; changes on the out-of-plane electric polarization due to in-plane strain have not been provided before. This study enriches the knowledge of the physical properties of two-dimensional group-IV monochalcogenides.

\section{Acknowledgments}
This work was funded by an Early Career Grant from the U.S. Department of Energy, Office of Basic Energy Sciences (Award  DE$-$SC0016139). Calculations were performed at Cori at NERSC, a U.S. DOE Office of Science User Facility operated under Contract No. DE$-$AC02$-$05CH11231.


\begin{thebibliography}{63}%
\makeatletter
\providecommand \@ifxundefined [1]{%
 \@ifx{#1\undefined}
}%
\providecommand \@ifnum [1]{%
 \ifnum #1\expandafter \@firstoftwo
 \else \expandafter \@secondoftwo
 \fi
}%
\providecommand \@ifx [1]{%
 \ifx #1\expandafter \@firstoftwo
 \else \expandafter \@secondoftwo
 \fi
}%
\providecommand \natexlab [1]{#1}%
\providecommand \enquote  [1]{``#1''}%
\providecommand \bibnamefont  [1]{#1}%
\providecommand \bibfnamefont [1]{#1}%
\providecommand \citenamefont [1]{#1}%
\providecommand \href@noop [0]{\@secondoftwo}%
\providecommand \href [0]{\begingroup \@sanitize@url \@href}%
\providecommand \@href[1]{\@@startlink{#1}\@@href}%
\providecommand \@@href[1]{\endgroup#1\@@endlink}%
\providecommand \@sanitize@url [0]{\catcode `\\12\catcode `\$12\catcode
  `\&12\catcode `\#12\catcode `\^12\catcode `\_12\catcode `\%12\relax}%
\providecommand \@@startlink[1]{}%
\providecommand \@@endlink[0]{}%
\providecommand \url  [0]{\begingroup\@sanitize@url \@url }%
\providecommand \@url [1]{\endgroup\@href {#1}{\urlprefix }}%
\providecommand \urlprefix  [0]{URL }%
\providecommand \Eprint [0]{\href }%
\providecommand \doibase [0]{http://dx.doi.org/}%
\providecommand \selectlanguage [0]{\@gobble}%
\providecommand \bibinfo  [0]{\@secondoftwo}%
\providecommand \bibfield  [0]{\@secondoftwo}%
\providecommand \translation [1]{[#1]}%
\providecommand \BibitemOpen [0]{}%
\providecommand \bibitemStop [0]{}%
\providecommand \bibitemNoStop [0]{.\EOS\space}%
\providecommand \EOS [0]{\spacefactor3000\relax}%
\providecommand \BibitemShut  [1]{\csname bibitem#1\endcsname}%
\let\auto@bib@innerbib\@empty
\bibitem [{\citenamefont {Mounet}\ \emph {et~al.}(2018)\citenamefont {Mounet},
  \citenamefont {Gibertini}, \citenamefont {Schwaller}, \citenamefont {Campi},
  \citenamefont {Merkys}, \citenamefont {Marrazzo}, \citenamefont {Sohier},
  \citenamefont {Castelli}, \citenamefont {Cepellotti}, \citenamefont {Pizzi},\
  and\ \citenamefont {Marzari}}]{marzari}%
  \BibitemOpen
  \bibfield  {author} {\bibinfo {author} {\bibfnamefont {N.}~\bibnamefont
  {Mounet}}, \bibinfo {author} {\bibfnamefont {M.}~\bibnamefont {Gibertini}},
  \bibinfo {author} {\bibfnamefont {P.}~\bibnamefont {Schwaller}}, \bibinfo
  {author} {\bibfnamefont {D.}~\bibnamefont {Campi}}, \bibinfo {author}
  {\bibfnamefont {A.}~\bibnamefont {Merkys}}, \bibinfo {author} {\bibfnamefont
  {A.}~\bibnamefont {Marrazzo}}, \bibinfo {author} {\bibfnamefont
  {T.}~\bibnamefont {Sohier}}, \bibinfo {author} {\bibfnamefont {I.~E.}\
  \bibnamefont {Castelli}}, \bibinfo {author} {\bibfnamefont {A.}~\bibnamefont
  {Cepellotti}}, \bibinfo {author} {\bibfnamefont {G.}~\bibnamefont {Pizzi}}, \
  and\ \bibinfo {author} {\bibfnamefont {N.}~\bibnamefont {Marzari}},\ }\href
  {\doibase 10.1038/s41565-017-0035-5} {\bibfield  {journal} {\bibinfo
  {journal} {Nat. Nanotech.}\ }\textbf {\bibinfo {volume} {13}},\ \bibinfo
  {pages} {246} (\bibinfo {year} {2018})}\BibitemShut {NoStop}%
\bibitem [{\citenamefont {Haastrup}\ \emph {et~al.}(2018)\citenamefont
  {Haastrup}, \citenamefont {Strange}, \citenamefont {Pandey}, \citenamefont
  {Deilmann}, \citenamefont {Schmidt}, \citenamefont {Hinsche}, \citenamefont
  {Gjerding}, \citenamefont {Torelli}, \citenamefont {Larsen}, \citenamefont
  {Riis-Jensen}, \citenamefont {Gath}, \citenamefont {Jacobsen}, \citenamefont
  {Mortensen}, \citenamefont {Olsen},\ and\ \citenamefont
  {Thygesen}}]{Haastrup_2018}%
  \BibitemOpen
  \bibfield  {author} {\bibinfo {author} {\bibfnamefont {S.}~\bibnamefont
  {Haastrup}}, \bibinfo {author} {\bibfnamefont {M.}~\bibnamefont {Strange}},
  \bibinfo {author} {\bibfnamefont {M.}~\bibnamefont {Pandey}}, \bibinfo
  {author} {\bibfnamefont {T.}~\bibnamefont {Deilmann}}, \bibinfo {author}
  {\bibfnamefont {P.~S.}\ \bibnamefont {Schmidt}}, \bibinfo {author}
  {\bibfnamefont {N.~F.}\ \bibnamefont {Hinsche}}, \bibinfo {author}
  {\bibfnamefont {M.~N.}\ \bibnamefont {Gjerding}}, \bibinfo {author}
  {\bibfnamefont {D.}~\bibnamefont {Torelli}}, \bibinfo {author} {\bibfnamefont
  {P.~M.}\ \bibnamefont {Larsen}}, \bibinfo {author} {\bibfnamefont {A.~C.}\
  \bibnamefont {Riis-Jensen}}, \bibinfo {author} {\bibfnamefont
  {J.}~\bibnamefont {Gath}}, \bibinfo {author} {\bibfnamefont {K.~W.}\
  \bibnamefont {Jacobsen}}, \bibinfo {author} {\bibfnamefont {J.~J.}\
  \bibnamefont {Mortensen}}, \bibinfo {author} {\bibfnamefont {T.}~\bibnamefont
  {Olsen}}, \ and\ \bibinfo {author} {\bibfnamefont {K.~S.}\ \bibnamefont
  {Thygesen}},\ }\href {\doibase 10.1088/2053-1583/aacfc1} {\bibfield
  {journal} {\bibinfo  {journal} {2D Materials}\ }\textbf {\bibinfo {volume}
  {5}},\ \bibinfo {pages} {042002} (\bibinfo {year} {2018})}\BibitemShut
  {NoStop}%
\bibitem [{\citenamefont {Tritsaris}\ \emph
  {et~al.}(2013{\natexlab{a}})\citenamefont {Tritsaris}, \citenamefont
  {Malone},\ and\ \citenamefont {Kaxiras}}]{tritsaris_jap_2013_sns}%
  \BibitemOpen
  \bibfield  {author} {\bibinfo {author} {\bibfnamefont {G.}~\bibnamefont
  {Tritsaris}}, \bibinfo {author} {\bibfnamefont {B.}~\bibnamefont {Malone}}, \
  and\ \bibinfo {author} {\bibfnamefont {E.}~\bibnamefont {Kaxiras}},\ }\href
  {\doibase 10.1063/1.4811455} {\bibfield  {journal} {\bibinfo  {journal} {J.
  Appl. Phys.}\ }\textbf {\bibinfo {volume} {113}},\ \bibinfo {pages} {233507}
  (\bibinfo {year} {2013}{\natexlab{a}})}\BibitemShut {NoStop}%
\bibitem [{\citenamefont {Singh}\ and\ \citenamefont
  {Hennig}(2014)}]{singh_apl_2014_ges_gese_sns_snse}%
  \BibitemOpen
  \bibfield  {author} {\bibinfo {author} {\bibfnamefont {A.~K.}\ \bibnamefont
  {Singh}}\ and\ \bibinfo {author} {\bibfnamefont {R.~G.}\ \bibnamefont
  {Hennig}},\ }\href {\doibase 10.1063/1.4891230} {\bibfield  {journal}
  {\bibinfo  {journal} {Appl. Phys. Lett.}\ }\textbf {\bibinfo {volume}
  {105}},\ \bibinfo {pages} {042103} (\bibinfo {year} {2014})}\BibitemShut
  {NoStop}%
\bibitem [{\citenamefont {Hu}\ and\ \citenamefont {Dong}(2016)}]{Hu2016}%
  \BibitemOpen
  \bibfield  {author} {\bibinfo {author} {\bibfnamefont {T.}~\bibnamefont
  {Hu}}\ and\ \bibinfo {author} {\bibfnamefont {J.}~\bibnamefont {Dong}},\
  }\href {\doibase 10.1039/C6CP06734D} {\bibfield  {journal} {\bibinfo
  {journal} {Phys. Chem. Chem. Phys.}\ }\textbf {\bibinfo {volume} {18}},\
  \bibinfo {pages} {32514} (\bibinfo {year} {2016})}\BibitemShut {NoStop}%
\bibitem [{\citenamefont {Ji}\ \emph {et~al.}(2017)\citenamefont {Ji},
  \citenamefont {Yang}, \citenamefont {Dong}, \citenamefont {Hou},
  \citenamefont {Wang},\ and\ \citenamefont {Li}}]{ref1}%
  \BibitemOpen
  \bibfield  {author} {\bibinfo {author} {\bibfnamefont {Y.}~\bibnamefont
  {Ji}}, \bibinfo {author} {\bibfnamefont {M.}~\bibnamefont {Yang}}, \bibinfo
  {author} {\bibfnamefont {H.}~\bibnamefont {Dong}}, \bibinfo {author}
  {\bibfnamefont {T.}~\bibnamefont {Hou}}, \bibinfo {author} {\bibfnamefont
  {L.}~\bibnamefont {Wang}}, \ and\ \bibinfo {author} {\bibfnamefont
  {Y.}~\bibnamefont {Li}},\ }\href {\doibase 10.1039/C7NR00688H} {\bibfield
  {journal} {\bibinfo  {journal} {Nanoscale}\ }\textbf {\bibinfo {volume}
  {9}},\ \bibinfo {pages} {8608} (\bibinfo {year} {2017})}\BibitemShut
  {NoStop}%
\bibitem [{\citenamefont {Gu}\ \emph {et~al.}(2019)\citenamefont {Gu},
  \citenamefont {Tao}, \citenamefont {Chen}, \citenamefont {Zhu}, \citenamefont
  {Ouyang},\ and\ \citenamefont {Peng}}]{Gu2019}%
  \BibitemOpen
  \bibfield  {author} {\bibinfo {author} {\bibfnamefont {D.}~\bibnamefont
  {Gu}}, \bibinfo {author} {\bibfnamefont {X.}~\bibnamefont {Tao}}, \bibinfo
  {author} {\bibfnamefont {H.}~\bibnamefont {Chen}}, \bibinfo {author}
  {\bibfnamefont {W.}~\bibnamefont {Zhu}}, \bibinfo {author} {\bibfnamefont
  {Y.}~\bibnamefont {Ouyang}}, \ and\ \bibinfo {author} {\bibfnamefont
  {Q.}~\bibnamefont {Peng}},\ }\href {\doibase 10.1039/C8NR08908F} {\bibfield
  {journal} {\bibinfo  {journal} {Nanoscale}\ }\textbf {\bibinfo {volume}
  {11}},\ \bibinfo {pages} {2335} (\bibinfo {year} {2019})}\BibitemShut
  {NoStop}%
\bibitem [{\citenamefont {Chang}\ \emph {et~al.}(2016)\citenamefont {Chang},
  \citenamefont {Liu}, \citenamefont {Lin}, \citenamefont {Wang}, \citenamefont
  {Zhao}, \citenamefont {Zhang}, \citenamefont {Jin}, \citenamefont {Zhong},
  \citenamefont {Hu}, \citenamefont {Duan}, \citenamefont {Zhang},
  \citenamefont {Fu}, \citenamefont {Xue}, \citenamefont {Chen},\ and\
  \citenamefont {Ji}}]{Chang2016}%
  \BibitemOpen
  \bibfield  {author} {\bibinfo {author} {\bibfnamefont {K.}~\bibnamefont
  {Chang}}, \bibinfo {author} {\bibfnamefont {J.}~\bibnamefont {Liu}}, \bibinfo
  {author} {\bibfnamefont {H.}~\bibnamefont {Lin}}, \bibinfo {author}
  {\bibfnamefont {N.}~\bibnamefont {Wang}}, \bibinfo {author} {\bibfnamefont
  {K.}~\bibnamefont {Zhao}}, \bibinfo {author} {\bibfnamefont {A.}~\bibnamefont
  {Zhang}}, \bibinfo {author} {\bibfnamefont {F.}~\bibnamefont {Jin}}, \bibinfo
  {author} {\bibfnamefont {Y.}~\bibnamefont {Zhong}}, \bibinfo {author}
  {\bibfnamefont {X.}~\bibnamefont {Hu}}, \bibinfo {author} {\bibfnamefont
  {W.}~\bibnamefont {Duan}}, \bibinfo {author} {\bibfnamefont {Q.}~\bibnamefont
  {Zhang}}, \bibinfo {author} {\bibfnamefont {L.}~\bibnamefont {Fu}}, \bibinfo
  {author} {\bibfnamefont {Q.-K.}\ \bibnamefont {Xue}}, \bibinfo {author}
  {\bibfnamefont {X.}~\bibnamefont {Chen}}, \ and\ \bibinfo {author}
  {\bibfnamefont {S.-H.}\ \bibnamefont {Ji}},\ }\href {\doibase
  10.1126/science.aad8609} {\bibfield  {journal} {\bibinfo  {journal}
  {Science}\ }\textbf {\bibinfo {volume} {353}},\ \bibinfo {pages} {274}
  (\bibinfo {year} {2016})}\BibitemShut {NoStop}%
\bibitem [{\citenamefont {Higashitarumizu}\ \emph {et~al.}(2020)\citenamefont
  {Higashitarumizu}, \citenamefont {Kawamoto}, \citenamefont {Lee},
  \citenamefont {Lin}, \citenamefont {Chu}, \citenamefont {Yonemori},
  \citenamefont {Nishimura}, \citenamefont {Wakabayashi}, \citenamefont
  {Chang},\ and\ \citenamefont {Nagashio}}]{Higashitarumizu20_NC_SnS}%
  \BibitemOpen
  \bibfield  {author} {\bibinfo {author} {\bibfnamefont {N.}~\bibnamefont
  {Higashitarumizu}}, \bibinfo {author} {\bibfnamefont {H.}~\bibnamefont
  {Kawamoto}}, \bibinfo {author} {\bibfnamefont {C.-J.}\ \bibnamefont {Lee}},
  \bibinfo {author} {\bibfnamefont {B.-H.}\ \bibnamefont {Lin}}, \bibinfo
  {author} {\bibfnamefont {F.-H.}\ \bibnamefont {Chu}}, \bibinfo {author}
  {\bibfnamefont {I.}~\bibnamefont {Yonemori}}, \bibinfo {author}
  {\bibfnamefont {T.}~\bibnamefont {Nishimura}}, \bibinfo {author}
  {\bibfnamefont {K.}~\bibnamefont {Wakabayashi}}, \bibinfo {author}
  {\bibfnamefont {W.-H.}\ \bibnamefont {Chang}}, \ and\ \bibinfo {author}
  {\bibfnamefont {K.}~\bibnamefont {Nagashio}},\ }\href {\doibase
  10.1038/s41467-020-16291-9} {\bibfield  {journal} {\bibinfo  {journal} {Nat.
  Commun.}\ }\textbf {\bibinfo {volume} {11}},\ \bibinfo {pages} {2428}
  (\bibinfo {year} {2020})}\BibitemShut {NoStop}%
\bibitem [{\citenamefont {Chang}\ \emph {et~al.}(2020)\citenamefont {Chang},
  \citenamefont {K{\"u}ster}, \citenamefont {Miller}, \citenamefont {Ji},
  \citenamefont {Zhang}, \citenamefont {Sessi}, \citenamefont {Barraza-Lopez},\
  and\ \citenamefont {Parkin}}]{Chang20_arxiv_SnSe}%
  \BibitemOpen
  \bibfield  {author} {\bibinfo {author} {\bibfnamefont {K.}~\bibnamefont
  {Chang}}, \bibinfo {author} {\bibfnamefont {F.}~\bibnamefont {K{\"u}ster}},
  \bibinfo {author} {\bibfnamefont {B.~J.}\ \bibnamefont {Miller}}, \bibinfo
  {author} {\bibfnamefont {J.-R.}\ \bibnamefont {Ji}}, \bibinfo {author}
  {\bibfnamefont {J.-L.}\ \bibnamefont {Zhang}}, \bibinfo {author}
  {\bibfnamefont {P.}~\bibnamefont {Sessi}}, \bibinfo {author} {\bibfnamefont
  {S.}~\bibnamefont {Barraza-Lopez}}, \ and\ \bibinfo {author} {\bibfnamefont
  {S.~S.~P.}\ \bibnamefont {Parkin}},\ }\href {http://arxiv.org/abs/2004.03884}
  {\bibfield  {journal} {\bibinfo  {journal} {Nano Lett. (accepted);
  arXiv:2004.03884}\ } (\bibinfo {year} {2020})}\BibitemShut {NoStop}%
\bibitem [{\citenamefont {Sutter}\ \emph {et~al.}(2019)\citenamefont {Sutter},
  \citenamefont {Wimer},\ and\ \citenamefont {Sutter}}]{sutterNature2019}%
  \BibitemOpen
  \bibfield  {author} {\bibinfo {author} {\bibfnamefont {P.}~\bibnamefont
  {Sutter}}, \bibinfo {author} {\bibfnamefont {S.}~\bibnamefont {Wimer}}, \
  and\ \bibinfo {author} {\bibfnamefont {E.}~\bibnamefont {Sutter}},\ }\href
  {\doibase 10.1038/s41586-019-1147-x} {\bibfield  {journal} {\bibinfo
  {journal} {Nature}\ }\textbf {\bibinfo {volume} {570}},\ \bibinfo {pages}
  {354} (\bibinfo {year} {2019})}\BibitemShut {NoStop}%
\bibitem [{\citenamefont {Tritsaris}\ \emph
  {et~al.}(2013{\natexlab{b}})\citenamefont {Tritsaris}, \citenamefont
  {Malone},\ and\ \citenamefont {Kaxiras}}]{kaxiras}%
  \BibitemOpen
  \bibfield  {author} {\bibinfo {author} {\bibfnamefont {G.~A.}\ \bibnamefont
  {Tritsaris}}, \bibinfo {author} {\bibfnamefont {B.~D.}\ \bibnamefont
  {Malone}}, \ and\ \bibinfo {author} {\bibfnamefont {E.}~\bibnamefont
  {Kaxiras}},\ }\href {\doibase 10.1063/1.4811455} {\bibfield  {journal}
  {\bibinfo  {journal} {J. Appl. Phys.}\ }\textbf {\bibinfo {volume} {113}},\
  \bibinfo {pages} {233507} (\bibinfo {year} {2013}{\natexlab{b}})}\BibitemShut
  {NoStop}%
\bibitem [{\citenamefont {Rodin}\ \emph {et~al.}(2016)\citenamefont {Rodin},
  \citenamefont {Gomes}, \citenamefont {Carvalho},\ and\ \citenamefont
  {Castro~Neto}}]{rodin_prb_2016_sns}%
  \BibitemOpen
  \bibfield  {author} {\bibinfo {author} {\bibfnamefont {A.~S.}\ \bibnamefont
  {Rodin}}, \bibinfo {author} {\bibfnamefont {L.~C.}\ \bibnamefont {Gomes}},
  \bibinfo {author} {\bibfnamefont {A.}~\bibnamefont {Carvalho}}, \ and\
  \bibinfo {author} {\bibfnamefont {A.~H.}\ \bibnamefont {Castro~Neto}},\
  }\href {\doibase 10.1103/PhysRevB.93.045431} {\bibfield  {journal} {\bibinfo
  {journal} {Phys. Rev. B}\ }\textbf {\bibinfo {volume} {93}},\ \bibinfo
  {pages} {045431} (\bibinfo {year} {2016})}\BibitemShut {NoStop}%
\bibitem [{\citenamefont {Yang}\ \emph {et~al.}(2016)\citenamefont {Yang},
  \citenamefont {Zhang}, \citenamefont {Yin}, \citenamefont {Gong},
  \citenamefont {Yakobson},\ and\ \citenamefont
  {Wei}}]{yang_nanolett_2015_sis}%
  \BibitemOpen
  \bibfield  {author} {\bibinfo {author} {\bibfnamefont {J.-H.}\ \bibnamefont
  {Yang}}, \bibinfo {author} {\bibfnamefont {Y.}~\bibnamefont {Zhang}},
  \bibinfo {author} {\bibfnamefont {W.-J.}\ \bibnamefont {Yin}}, \bibinfo
  {author} {\bibfnamefont {X.~G.}\ \bibnamefont {Gong}}, \bibinfo {author}
  {\bibfnamefont {B.~I.}\ \bibnamefont {Yakobson}}, \ and\ \bibinfo {author}
  {\bibfnamefont {S.-H.}\ \bibnamefont {Wei}},\ }\href {\doibase
  10.1021/acs.nanolett.5b04341} {\bibfield  {journal} {\bibinfo  {journal}
  {Nano Lett.}\ }\textbf {\bibinfo {volume} {16}},\ \bibinfo {pages} {1110}
  (\bibinfo {year} {2016})}\BibitemShut {NoStop}%
\bibitem [{\citenamefont {Qiao}\ \emph {et~al.}(2018)\citenamefont {Qiao},
  \citenamefont {Chen}, \citenamefont {Wang},\ and\ \citenamefont {Li}}]{GeTe}%
  \BibitemOpen
  \bibfield  {author} {\bibinfo {author} {\bibfnamefont {M.}~\bibnamefont
  {Qiao}}, \bibinfo {author} {\bibfnamefont {Y.}~\bibnamefont {Chen}}, \bibinfo
  {author} {\bibfnamefont {Y.}~\bibnamefont {Wang}}, \ and\ \bibinfo {author}
  {\bibfnamefont {Y.}~\bibnamefont {Li}},\ }\href {\doibase 10.1039/C7TA10360C}
  {\bibfield  {journal} {\bibinfo  {journal} {J. Mater. Chem. A}\ }\textbf
  {\bibinfo {volume} {6}},\ \bibinfo {pages} {4119} (\bibinfo {year}
  {2018})}\BibitemShut {NoStop}%
\bibitem [{\citenamefont {Zhu}\ and\ \citenamefont
  {Tom\'anek}(2014)}]{Zhu2014}%
  \BibitemOpen
  \bibfield  {author} {\bibinfo {author} {\bibfnamefont {Z.}~\bibnamefont
  {Zhu}}\ and\ \bibinfo {author} {\bibfnamefont {D.}~\bibnamefont
  {Tom\'anek}},\ }\href {\doibase 10.1103/PhysRevLett.112.176802} {\bibfield
  {journal} {\bibinfo  {journal} {Phys. Rev. Lett.}\ }\textbf {\bibinfo
  {volume} {112}},\ \bibinfo {pages} {176802} (\bibinfo {year}
  {2014})}\BibitemShut {NoStop}%
\bibitem [{\citenamefont {Rivero}\ \emph {et~al.}(2014)\citenamefont {Rivero},
  \citenamefont {Yan}, \citenamefont {Garc\'{\i}a-Su\'arez}, \citenamefont
  {Ferrer},\ and\ \citenamefont {Barraza-Lopez}}]{Rivero2014}%
  \BibitemOpen
  \bibfield  {author} {\bibinfo {author} {\bibfnamefont {P.}~\bibnamefont
  {Rivero}}, \bibinfo {author} {\bibfnamefont {J.-A.}\ \bibnamefont {Yan}},
  \bibinfo {author} {\bibfnamefont {V.~M.}\ \bibnamefont
  {Garc\'{\i}a-Su\'arez}}, \bibinfo {author} {\bibfnamefont {J.}~\bibnamefont
  {Ferrer}}, \ and\ \bibinfo {author} {\bibfnamefont {S.}~\bibnamefont
  {Barraza-Lopez}},\ }\href {\doibase 10.1103/PhysRevB.90.241408} {\bibfield
  {journal} {\bibinfo  {journal} {Phys. Rev. B}\ }\textbf {\bibinfo {volume}
  {90}},\ \bibinfo {pages} {241408} (\bibinfo {year} {2014})}\BibitemShut
  {NoStop}%
\bibitem [{\citenamefont {Cahangirov}\ \emph {et~al.}(2009)\citenamefont
  {Cahangirov}, \citenamefont {Topsakal}, \citenamefont {Akt\"urk},
  \citenamefont {\ifmmode~\mbox{\c{S}}\else \c{S}\fi{}ahin},\ and\
  \citenamefont {Ciraci}}]{Cahangirov2009}%
  \BibitemOpen
  \bibfield  {author} {\bibinfo {author} {\bibfnamefont {S.}~\bibnamefont
  {Cahangirov}}, \bibinfo {author} {\bibfnamefont {M.}~\bibnamefont
  {Topsakal}}, \bibinfo {author} {\bibfnamefont {E.}~\bibnamefont {Akt\"urk}},
  \bibinfo {author} {\bibfnamefont {H.}~\bibnamefont
  {\ifmmode~\mbox{\c{S}}\else \c{S}\fi{}ahin}}, \ and\ \bibinfo {author}
  {\bibfnamefont {S.}~\bibnamefont {Ciraci}},\ }\href {\doibase
  10.1103/PhysRevLett.102.236804} {\bibfield  {journal} {\bibinfo  {journal}
  {Phys. Rev. Lett.}\ }\textbf {\bibinfo {volume} {102}},\ \bibinfo {pages}
  {236804} (\bibinfo {year} {2009})}\BibitemShut {NoStop}%
\bibitem [{\citenamefont {Xu}\ \emph {et~al.}(2013)\citenamefont {Xu},
  \citenamefont {Yan}, \citenamefont {Zhang}, \citenamefont {Wang},
  \citenamefont {Xu}, \citenamefont {Tang}, \citenamefont {Duan},\ and\
  \citenamefont {Zhang}}]{PhysRevLett.111.136804}%
  \BibitemOpen
  \bibfield  {author} {\bibinfo {author} {\bibfnamefont {Y.}~\bibnamefont
  {Xu}}, \bibinfo {author} {\bibfnamefont {B.}~\bibnamefont {Yan}}, \bibinfo
  {author} {\bibfnamefont {H.-J.}\ \bibnamefont {Zhang}}, \bibinfo {author}
  {\bibfnamefont {J.}~\bibnamefont {Wang}}, \bibinfo {author} {\bibfnamefont
  {G.}~\bibnamefont {Xu}}, \bibinfo {author} {\bibfnamefont {P.}~\bibnamefont
  {Tang}}, \bibinfo {author} {\bibfnamefont {W.}~\bibnamefont {Duan}}, \ and\
  \bibinfo {author} {\bibfnamefont {S.-C.}\ \bibnamefont {Zhang}},\ }\href
  {\doibase 10.1103/PhysRevLett.111.136804} {\bibfield  {journal} {\bibinfo
  {journal} {Phys. Rev. Lett.}\ }\textbf {\bibinfo {volume} {111}},\ \bibinfo
  {pages} {136804} (\bibinfo {year} {2013})}\BibitemShut {NoStop}%
\bibitem [{\citenamefont {Duerloo}\ \emph {et~al.}(2012)\citenamefont
  {Duerloo}, \citenamefont {Ong},\ and\ \citenamefont {Reed}}]{duerloo}%
  \BibitemOpen
  \bibfield  {author} {\bibinfo {author} {\bibfnamefont {K.-A.~N.}\
  \bibnamefont {Duerloo}}, \bibinfo {author} {\bibfnamefont {M.~T.}\
  \bibnamefont {Ong}}, \ and\ \bibinfo {author} {\bibfnamefont {E.~J.}\
  \bibnamefont {Reed}},\ }\href {\doibase 10.1021/jz3012436} {\bibfield
  {journal} {\bibinfo  {journal} {J. Phys. Chem. Lett.}\ }\textbf {\bibinfo
  {volume} {3}},\ \bibinfo {pages} {2871} (\bibinfo {year} {2012})}\BibitemShut
  {NoStop}%
\bibitem [{\citenamefont {Wu}\ \emph {et~al.}(2014)\citenamefont {Wu},
  \citenamefont {Wang}, \citenamefont {Li}, \citenamefont {Zhang},
  \citenamefont {Lin}, \citenamefont {Niu}, \citenamefont {Chenet},
  \citenamefont {Zhang}, \citenamefont {Hao}, \citenamefont {Heinz},
  \citenamefont {Hone},\ and\ \citenamefont {Wang}}]{piezoMoS2}%
  \BibitemOpen
  \bibfield  {author} {\bibinfo {author} {\bibfnamefont {W.}~\bibnamefont
  {Wu}}, \bibinfo {author} {\bibfnamefont {L.}~\bibnamefont {Wang}}, \bibinfo
  {author} {\bibfnamefont {Y.}~\bibnamefont {Li}}, \bibinfo {author}
  {\bibfnamefont {F.}~\bibnamefont {Zhang}}, \bibinfo {author} {\bibfnamefont
  {L.}~\bibnamefont {Lin}}, \bibinfo {author} {\bibfnamefont {S.}~\bibnamefont
  {Niu}}, \bibinfo {author} {\bibfnamefont {D.}~\bibnamefont {Chenet}},
  \bibinfo {author} {\bibfnamefont {X.}~\bibnamefont {Zhang}}, \bibinfo
  {author} {\bibfnamefont {Y.}~\bibnamefont {Hao}}, \bibinfo {author}
  {\bibfnamefont {T.~F.}\ \bibnamefont {Heinz}}, \bibinfo {author}
  {\bibfnamefont {J.}~\bibnamefont {Hone}}, \ and\ \bibinfo {author}
  {\bibfnamefont {Z.~L.}\ \bibnamefont {Wang}},\ }\href {\doibase
  10.1038/nature13792} {\bibfield  {journal} {\bibinfo  {journal} {Nature}\
  }\textbf {\bibinfo {volume} {514}},\ \bibinfo {pages} {470} (\bibinfo {year}
  {2014})}\BibitemShut {NoStop}%
\bibitem [{\citenamefont {Ares}\ \emph {et~al.}(2020)\citenamefont {Ares},
  \citenamefont {Cea}, \citenamefont {Holwill}, \citenamefont {Wang},
  \citenamefont {Roldán}, \citenamefont {Guinea}, \citenamefont {Andreeva},
  \citenamefont {Fumagalli}, \citenamefont {Novoselov},\ and\ \citenamefont
  {Woods}}]{piezohBN}%
  \BibitemOpen
  \bibfield  {author} {\bibinfo {author} {\bibfnamefont {P.}~\bibnamefont
  {Ares}}, \bibinfo {author} {\bibfnamefont {T.}~\bibnamefont {Cea}}, \bibinfo
  {author} {\bibfnamefont {M.}~\bibnamefont {Holwill}}, \bibinfo {author}
  {\bibfnamefont {Y.~B.}\ \bibnamefont {Wang}}, \bibinfo {author}
  {\bibfnamefont {R.}~\bibnamefont {Roldán}}, \bibinfo {author} {\bibfnamefont
  {F.}~\bibnamefont {Guinea}}, \bibinfo {author} {\bibfnamefont {D.~V.}\
  \bibnamefont {Andreeva}}, \bibinfo {author} {\bibfnamefont {L.}~\bibnamefont
  {Fumagalli}}, \bibinfo {author} {\bibfnamefont {K.~S.}\ \bibnamefont
  {Novoselov}}, \ and\ \bibinfo {author} {\bibfnamefont {C.~R.}\ \bibnamefont
  {Woods}},\ }\href {\doibase 10.1002/adma.201905504} {\bibfield  {journal}
  {\bibinfo  {journal} {Adv. Mater.}\ }\textbf {\bibinfo {volume} {32}},\
  \bibinfo {pages} {1905504} (\bibinfo {year} {2020})}\BibitemShut {NoStop}%
\bibitem [{\citenamefont {Toledo-Mar{\'\i}n}\ and\ \citenamefont
  {Naumis}(2017)}]{toledo}%
  \BibitemOpen
  \bibfield  {author} {\bibinfo {author} {\bibfnamefont {J.~Q.}\ \bibnamefont
  {Toledo-Mar{\'\i}n}}\ and\ \bibinfo {author} {\bibfnamefont {G.~G.}\
  \bibnamefont {Naumis}},\ }\href {\doibase 10.1063/1.4977517} {\bibfield
  {journal} {\bibinfo  {journal} {J. Chem. Phys.}\ }\textbf {\bibinfo {volume}
  {146}},\ \bibinfo {pages} {094506} (\bibinfo {year} {2017})}\BibitemShut
  {NoStop}%
\bibitem [{\citenamefont {Pick}\ \emph {et~al.}(1970)\citenamefont {Pick},
  \citenamefont {Cohen},\ and\ \citenamefont {Martin}}]{PhysRevB.1.910}%
  \BibitemOpen
  \bibfield  {author} {\bibinfo {author} {\bibfnamefont {R.~M.}\ \bibnamefont
  {Pick}}, \bibinfo {author} {\bibfnamefont {M.~H.}\ \bibnamefont {Cohen}}, \
  and\ \bibinfo {author} {\bibfnamefont {R.~M.}\ \bibnamefont {Martin}},\
  }\href {\doibase 10.1103/PhysRevB.1.910} {\bibfield  {journal} {\bibinfo
  {journal} {Phys. Rev. B}\ }\textbf {\bibinfo {volume} {1}},\ \bibinfo {pages}
  {910} (\bibinfo {year} {1970})}\BibitemShut {NoStop}%
\bibitem [{\citenamefont {Kresse}\ and\ \citenamefont
  {Furthm\"uller}(1996)}]{vasp}%
  \BibitemOpen
  \bibfield  {author} {\bibinfo {author} {\bibfnamefont {G.}~\bibnamefont
  {Kresse}}\ and\ \bibinfo {author} {\bibfnamefont {J.}~\bibnamefont
  {Furthm\"uller}},\ }\href {\doibase 10.1103/PhysRevB.54.11169} {\bibfield
  {journal} {\bibinfo  {journal} {Phys. Rev. B}\ }\textbf {\bibinfo {volume}
  {54}},\ \bibinfo {pages} {11169} (\bibinfo {year} {1996})}\BibitemShut
  {NoStop}%
\bibitem [{\citenamefont {Kresse}\ and\ \citenamefont {Joubert}(1999)}]{vasp1}%
  \BibitemOpen
  \bibfield  {author} {\bibinfo {author} {\bibfnamefont {G.}~\bibnamefont
  {Kresse}}\ and\ \bibinfo {author} {\bibfnamefont {D.}~\bibnamefont
  {Joubert}},\ }\href {\doibase 10.1103/PhysRevB.59.1758} {\bibfield  {journal}
  {\bibinfo  {journal} {Phys. Rev. B}\ }\textbf {\bibinfo {volume} {59}},\
  \bibinfo {pages} {1758} (\bibinfo {year} {1999})}\BibitemShut {NoStop}%
\bibitem [{\citenamefont {Berland}\ \emph {et~al.}(2015)\citenamefont
  {Berland}, \citenamefont {Cooper}, \citenamefont {Lee}, \citenamefont
  {Schr{\"o}der}, \citenamefont {Thonhauser}, \citenamefont {Hyldgaard},\ and\
  \citenamefont {Lundqvist}}]{reviewvdw}%
  \BibitemOpen
  \bibfield  {author} {\bibinfo {author} {\bibfnamefont {K.}~\bibnamefont
  {Berland}}, \bibinfo {author} {\bibfnamefont {V.~R.}\ \bibnamefont {Cooper}},
  \bibinfo {author} {\bibfnamefont {K.}~\bibnamefont {Lee}}, \bibinfo {author}
  {\bibfnamefont {E.}~\bibnamefont {Schr{\"o}der}}, \bibinfo {author}
  {\bibfnamefont {T.}~\bibnamefont {Thonhauser}}, \bibinfo {author}
  {\bibfnamefont {P.}~\bibnamefont {Hyldgaard}}, \ and\ \bibinfo {author}
  {\bibfnamefont {B.~I.}\ \bibnamefont {Lundqvist}},\ }\href {\doibase
  10.1088/0034-4885/78/6/066501} {\bibfield  {journal} {\bibinfo  {journal}
  {Rep. Prog. Phys.}\ }\textbf {\bibinfo {volume} {78}},\ \bibinfo {pages}
  {066501} (\bibinfo {year} {2015})}\BibitemShut {NoStop}%
\bibitem [{\citenamefont {Klime{\v{s}}}\ \emph {et~al.}(2009)\citenamefont
  {Klime{\v{s}}}, \citenamefont {Bowler},\ and\ \citenamefont
  {Michaelides}}]{klimes1}%
  \BibitemOpen
  \bibfield  {author} {\bibinfo {author} {\bibfnamefont {J.}~\bibnamefont
  {Klime{\v{s}}}}, \bibinfo {author} {\bibfnamefont {D.~R.}\ \bibnamefont
  {Bowler}}, \ and\ \bibinfo {author} {\bibfnamefont {A.}~\bibnamefont
  {Michaelides}},\ }\href {\doibase 10.1088/0953-8984/22/2/022201} {\bibfield
  {journal} {\bibinfo  {journal} {J. Phys.: Condens. Matter}\ }\textbf
  {\bibinfo {volume} {22}},\ \bibinfo {pages} {022201} (\bibinfo {year}
  {2009})}\BibitemShut {NoStop}%
\bibitem [{\citenamefont {Lee}\ \emph {et~al.}(2010)\citenamefont {Lee},
  \citenamefont {Murray}, \citenamefont {Kong}, \citenamefont {Lundqvist},\
  and\ \citenamefont {Langreth}}]{klimes2}%
  \BibitemOpen
  \bibfield  {author} {\bibinfo {author} {\bibfnamefont {K.}~\bibnamefont
  {Lee}}, \bibinfo {author} {\bibfnamefont {E.~D.}\ \bibnamefont {Murray}},
  \bibinfo {author} {\bibfnamefont {L.}~\bibnamefont {Kong}}, \bibinfo {author}
  {\bibfnamefont {B.~I.}\ \bibnamefont {Lundqvist}}, \ and\ \bibinfo {author}
  {\bibfnamefont {D.~C.}\ \bibnamefont {Langreth}},\ }\href {\doibase
  10.1103/PhysRevB.82.081101} {\bibfield  {journal} {\bibinfo  {journal} {Phys.
  Rev. B}\ }\textbf {\bibinfo {volume} {82}},\ \bibinfo {pages} {081101}
  (\bibinfo {year} {2010})}\BibitemShut {NoStop}%
\bibitem [{\citenamefont {Klime\ifmmode~\check{s}\else \v{s}\fi{}}\ \emph
  {et~al.}(2011)\citenamefont {Klime\ifmmode~\check{s}\else \v{s}\fi{}},
  \citenamefont {Bowler},\ and\ \citenamefont {Michaelides}}]{klimes3}%
  \BibitemOpen
  \bibfield  {author} {\bibinfo {author} {\bibfnamefont {J.~c.~v.}\
  \bibnamefont {Klime\ifmmode~\check{s}\else \v{s}\fi{}}}, \bibinfo {author}
  {\bibfnamefont {D.~R.}\ \bibnamefont {Bowler}}, \ and\ \bibinfo {author}
  {\bibfnamefont {A.}~\bibnamefont {Michaelides}},\ }\href {\doibase
  10.1103/PhysRevB.83.195131} {\bibfield  {journal} {\bibinfo  {journal} {Phys.
  Rev. B}\ }\textbf {\bibinfo {volume} {83}},\ \bibinfo {pages} {195131}
  (\bibinfo {year} {2011})}\BibitemShut {NoStop}%
\bibitem [{\citenamefont {Togo}\ and\ \citenamefont {Tanaka}(2015)}]{phonophy}%
  \BibitemOpen
  \bibfield  {author} {\bibinfo {author} {\bibfnamefont {A.}~\bibnamefont
  {Togo}}\ and\ \bibinfo {author} {\bibfnamefont {I.}~\bibnamefont {Tanaka}},\
  }\href {\doibase https://doi.org/10.1016/j.scriptamat.2015.07.021} {\bibfield
   {journal} {\bibinfo  {journal} {Scr. Mater.}\ }\textbf {\bibinfo {volume}
  {108}},\ \bibinfo {pages} {1} (\bibinfo {year} {2015})}\BibitemShut {NoStop}%
\bibitem [{\citenamefont {Resta}\ and\ \citenamefont
  {Vanderbilt}(2007)}]{modern}%
  \BibitemOpen
  \bibfield  {author} {\bibinfo {author} {\bibfnamefont {R.}~\bibnamefont
  {Resta}}\ and\ \bibinfo {author} {\bibfnamefont {D.}~\bibnamefont
  {Vanderbilt}},\ }\enquote {\bibinfo {title} {Theory of polarization: A modern
  approach},}\ \ (\bibinfo  {publisher} {Springer},\ \bibinfo {address}
  {Berlin},\ \bibinfo {year} {2007})\ Chap.~\bibinfo {chapter} {2}, pp.\
  \bibinfo {pages} {31--68},\ \bibinfo {edition} {1st}\ ed.\BibitemShut {Stop}%
\bibitem [{\citenamefont {Car}\ and\ \citenamefont
  {Parrinello}(1985)}]{Car-Parrinello}%
  \BibitemOpen
  \bibfield  {author} {\bibinfo {author} {\bibfnamefont {R.}~\bibnamefont
  {Car}}\ and\ \bibinfo {author} {\bibfnamefont {M.}~\bibnamefont
  {Parrinello}},\ }\href {\doibase 10.1103/PhysRevLett.55.2471} {\bibfield
  {journal} {\bibinfo  {journal} {Phys. Rev. Lett.}\ }\textbf {\bibinfo
  {volume} {55}},\ \bibinfo {pages} {2471} (\bibinfo {year}
  {1985})}\BibitemShut {NoStop}%
\bibitem [{\citenamefont {Soler}\ \emph {et~al.}(2002)\citenamefont {Soler},
  \citenamefont {Artacho}, \citenamefont {Gale}, \citenamefont {Garc{\'{\i}}a},
  \citenamefont {Junquera}, \citenamefont {Ordej{\'{o}}n},\ and\ \citenamefont
  {S{\'{a}}nchez-Portal}}]{siesta}%
  \BibitemOpen
  \bibfield  {author} {\bibinfo {author} {\bibfnamefont {J.~M.}\ \bibnamefont
  {Soler}}, \bibinfo {author} {\bibfnamefont {E.}~\bibnamefont {Artacho}},
  \bibinfo {author} {\bibfnamefont {J.~D.}\ \bibnamefont {Gale}}, \bibinfo
  {author} {\bibfnamefont {A.}~\bibnamefont {Garc{\'{\i}}a}}, \bibinfo {author}
  {\bibfnamefont {J.}~\bibnamefont {Junquera}}, \bibinfo {author}
  {\bibfnamefont {P.}~\bibnamefont {Ordej{\'{o}}n}}, \ and\ \bibinfo {author}
  {\bibfnamefont {D.}~\bibnamefont {S{\'{a}}nchez-Portal}},\ }\href {\doibase
  10.1088/0953-8984/14/11/302} {\bibfield  {journal} {\bibinfo  {journal} {J.
  Phys.: Condens. Matter}\ }\textbf {\bibinfo {volume} {14}},\ \bibinfo {pages}
  {2745} (\bibinfo {year} {2002})}\BibitemShut {NoStop}%
\bibitem [{\citenamefont {Junquera}\ \emph {et~al.}(2001)\citenamefont
  {Junquera}, \citenamefont {Paz}, \citenamefont {S\'anchez-Portal},\ and\
  \citenamefont {Artacho}}]{Junquera2001}%
  \BibitemOpen
  \bibfield  {author} {\bibinfo {author} {\bibfnamefont {J.}~\bibnamefont
  {Junquera}}, \bibinfo {author} {\bibfnamefont {O.}~\bibnamefont {Paz}},
  \bibinfo {author} {\bibfnamefont {D.}~\bibnamefont {S\'anchez-Portal}}, \
  and\ \bibinfo {author} {\bibfnamefont {E.}~\bibnamefont {Artacho}},\ }\href
  {\doibase 10.1103/PhysRevB.64.235111} {\bibfield  {journal} {\bibinfo
  {journal} {Phys. Rev. B}\ }\textbf {\bibinfo {volume} {64}},\ \bibinfo
  {pages} {235111} (\bibinfo {year} {2001})}\BibitemShut {NoStop}%
\bibitem [{\citenamefont {Troullier}\ and\ \citenamefont
  {Martins}(1991)}]{Troullier}%
  \BibitemOpen
  \bibfield  {author} {\bibinfo {author} {\bibfnamefont {N.}~\bibnamefont
  {Troullier}}\ and\ \bibinfo {author} {\bibfnamefont {J.~L.}\ \bibnamefont
  {Martins}},\ }\href {\doibase 10.1103/PhysRevB.43.1993} {\bibfield  {journal}
  {\bibinfo  {journal} {Phys. Rev. B}\ }\textbf {\bibinfo {volume} {43}},\
  \bibinfo {pages} {1993} (\bibinfo {year} {1991})}\BibitemShut {NoStop}%
\bibitem [{\citenamefont {Rivero}\ \emph {et~al.}(2015)\citenamefont {Rivero},
  \citenamefont {Garcia-Suarez}, \citenamefont {Pere{\~n}iguez}, \citenamefont
  {Utt}, \citenamefont {Yang}, \citenamefont {Bellaiche}, \citenamefont {Park},
  \citenamefont {Ferrer},\ and\ \citenamefont {Barraza-Lopez}}]{rivero}%
  \BibitemOpen
  \bibfield  {author} {\bibinfo {author} {\bibfnamefont {P.}~\bibnamefont
  {Rivero}}, \bibinfo {author} {\bibfnamefont {V.~M.}\ \bibnamefont
  {Garcia-Suarez}}, \bibinfo {author} {\bibfnamefont {D.}~\bibnamefont
  {Pere{\~n}iguez}}, \bibinfo {author} {\bibfnamefont {K.}~\bibnamefont {Utt}},
  \bibinfo {author} {\bibfnamefont {Y.}~\bibnamefont {Yang}}, \bibinfo {author}
  {\bibfnamefont {L.}~\bibnamefont {Bellaiche}}, \bibinfo {author}
  {\bibfnamefont {K.}~\bibnamefont {Park}}, \bibinfo {author} {\bibfnamefont
  {J.}~\bibnamefont {Ferrer}}, \ and\ \bibinfo {author} {\bibfnamefont
  {S.}~\bibnamefont {Barraza-Lopez}},\ }\href {\doibase
  10.1016/j.commatsci.2014.11.026} {\bibfield  {journal} {\bibinfo  {journal}
  {Comp. Mat. Sci.}\ }\textbf {\bibinfo {volume} {98}},\ \bibinfo {pages} {372}
  (\bibinfo {year} {2015})}\BibitemShut {NoStop}%
\bibitem [{\citenamefont {Hyldgaard}\ \emph {et~al.}(2014)\citenamefont
  {Hyldgaard}, \citenamefont {Berland},\ and\ \citenamefont
  {Schr\"oder}}]{Hyldgaard}%
  \BibitemOpen
  \bibfield  {author} {\bibinfo {author} {\bibfnamefont {P.}~\bibnamefont
  {Hyldgaard}}, \bibinfo {author} {\bibfnamefont {K.}~\bibnamefont {Berland}},
  \ and\ \bibinfo {author} {\bibfnamefont {E.}~\bibnamefont {Schr\"oder}},\
  }\href {\doibase 10.1103/PhysRevB.90.075148} {\bibfield  {journal} {\bibinfo
  {journal} {Phys. Rev. B}\ }\textbf {\bibinfo {volume} {90}},\ \bibinfo
  {pages} {075148} (\bibinfo {year} {2014})}\BibitemShut {NoStop}%
\bibitem [{\citenamefont {Rom\'an-P\'erez}\ and\ \citenamefont
  {Soler}(2009)}]{soler}%
  \BibitemOpen
  \bibfield  {author} {\bibinfo {author} {\bibfnamefont {G.}~\bibnamefont
  {Rom\'an-P\'erez}}\ and\ \bibinfo {author} {\bibfnamefont {J.~M.}\
  \bibnamefont {Soler}},\ }\href {\doibase 10.1103/PhysRevLett.103.096102}
  {\bibfield  {journal} {\bibinfo  {journal} {Phys. Rev. Lett.}\ }\textbf
  {\bibinfo {volume} {103}},\ \bibinfo {pages} {096102} (\bibinfo {year}
  {2009})}\BibitemShut {NoStop}%
\bibitem [{\citenamefont {Mehboudi}\ \emph
  {et~al.}(2016{\natexlab{a}})\citenamefont {Mehboudi}, \citenamefont {Dorio},
  \citenamefont {Zhu}, \citenamefont {van~der Zande}, \citenamefont
  {Churchill}, \citenamefont {Pacheco-Sanjuan}, \citenamefont {Harriss},
  \citenamefont {Kumar},\ and\ \citenamefont {Barraza-Lopez}}]{Mehboudi-nl}%
  \BibitemOpen
  \bibfield  {author} {\bibinfo {author} {\bibfnamefont {M.}~\bibnamefont
  {Mehboudi}}, \bibinfo {author} {\bibfnamefont {A.~M.}\ \bibnamefont {Dorio}},
  \bibinfo {author} {\bibfnamefont {W.}~\bibnamefont {Zhu}}, \bibinfo {author}
  {\bibfnamefont {A.}~\bibnamefont {van~der Zande}}, \bibinfo {author}
  {\bibfnamefont {H.~O.~H.}\ \bibnamefont {Churchill}}, \bibinfo {author}
  {\bibfnamefont {A.~A.}\ \bibnamefont {Pacheco-Sanjuan}}, \bibinfo {author}
  {\bibfnamefont {E.~O.}\ \bibnamefont {Harriss}}, \bibinfo {author}
  {\bibfnamefont {P.}~\bibnamefont {Kumar}}, \ and\ \bibinfo {author}
  {\bibfnamefont {S.}~\bibnamefont {Barraza-Lopez}},\ }\href {\doibase
  10.1021/acs.nanolett.5b04613} {\bibfield  {journal} {\bibinfo  {journal}
  {Nano Lett.}\ }\textbf {\bibinfo {volume} {16}},\ \bibinfo {pages} {1704}
  (\bibinfo {year} {2016}{\natexlab{a}})}\BibitemShut {NoStop}%
\bibitem [{\citenamefont {Mehboudi}\ \emph
  {et~al.}(2016{\natexlab{b}})\citenamefont {Mehboudi}, \citenamefont
  {Fregoso}, \citenamefont {Yang}, \citenamefont {Zhu}, \citenamefont {van~der
  Zande}, \citenamefont {Ferrer}, \citenamefont {Bellaiche}, \citenamefont
  {Kumar},\ and\ \citenamefont {Barraza-Lopez}}]{Mehboudi-prl}%
  \BibitemOpen
  \bibfield  {author} {\bibinfo {author} {\bibfnamefont {M.}~\bibnamefont
  {Mehboudi}}, \bibinfo {author} {\bibfnamefont {B.~M.}\ \bibnamefont
  {Fregoso}}, \bibinfo {author} {\bibfnamefont {Y.}~\bibnamefont {Yang}},
  \bibinfo {author} {\bibfnamefont {W.}~\bibnamefont {Zhu}}, \bibinfo {author}
  {\bibfnamefont {A.}~\bibnamefont {van~der Zande}}, \bibinfo {author}
  {\bibfnamefont {J.}~\bibnamefont {Ferrer}}, \bibinfo {author} {\bibfnamefont
  {L.}~\bibnamefont {Bellaiche}}, \bibinfo {author} {\bibfnamefont
  {P.}~\bibnamefont {Kumar}}, \ and\ \bibinfo {author} {\bibfnamefont
  {S.}~\bibnamefont {Barraza-Lopez}},\ }\href {\doibase
  10.1103/PhysRevLett.117.246802} {\bibfield  {journal} {\bibinfo  {journal}
  {Phys. Rev. Lett.}\ }\textbf {\bibinfo {volume} {117}},\ \bibinfo {pages}
  {246802} (\bibinfo {year} {2016}{\natexlab{b}})}\BibitemShut {NoStop}%
\bibitem [{\citenamefont {Barraza-Lopez}\ \emph {et~al.}(2018)\citenamefont
  {Barraza-Lopez}, \citenamefont {Kaloni}, \citenamefont {Poudel},\ and\
  \citenamefont {Kumar}}]{Salvador2018-prb}%
  \BibitemOpen
  \bibfield  {author} {\bibinfo {author} {\bibfnamefont {S.}~\bibnamefont
  {Barraza-Lopez}}, \bibinfo {author} {\bibfnamefont {T.~P.}\ \bibnamefont
  {Kaloni}}, \bibinfo {author} {\bibfnamefont {S.~P.}\ \bibnamefont {Poudel}},
  \ and\ \bibinfo {author} {\bibfnamefont {P.}~\bibnamefont {Kumar}},\ }\href
  {\doibase 10.1103/PhysRevB.97.024110} {\bibfield  {journal} {\bibinfo
  {journal} {Phys. Rev. B}\ }\textbf {\bibinfo {volume} {97}},\ \bibinfo
  {pages} {024110} (\bibinfo {year} {2018})}\BibitemShut {NoStop}%
\bibitem [{\citenamefont {Mills}\ and\ \citenamefont
  {J\'onsson}(1994)}]{nudged1}%
  \BibitemOpen
  \bibfield  {author} {\bibinfo {author} {\bibfnamefont {G.}~\bibnamefont
  {Mills}}\ and\ \bibinfo {author} {\bibfnamefont {H.}~\bibnamefont
  {J\'onsson}},\ }\href {\doibase 10.1103/PhysRevLett.72.1124} {\bibfield
  {journal} {\bibinfo  {journal} {Phys. Rev. Lett.}\ }\textbf {\bibinfo
  {volume} {72}},\ \bibinfo {pages} {1124} (\bibinfo {year}
  {1994})}\BibitemShut {NoStop}%
\bibitem [{\citenamefont {Henkelman}\ \emph {et~al.}(2000)\citenamefont
  {Henkelman}, \citenamefont {Uberuaga},\ and\ \citenamefont
  {Jónsson}}]{nudged2}%
  \BibitemOpen
  \bibfield  {author} {\bibinfo {author} {\bibfnamefont {G.}~\bibnamefont
  {Henkelman}}, \bibinfo {author} {\bibfnamefont {B.~P.}\ \bibnamefont
  {Uberuaga}}, \ and\ \bibinfo {author} {\bibfnamefont {H.}~\bibnamefont
  {Jónsson}},\ }\href {\doibase 10.1063/1.1329672} {\bibfield  {journal}
  {\bibinfo  {journal} {J. Chem. Phys.}\ }\textbf {\bibinfo {volume} {113}},\
  \bibinfo {pages} {9901} (\bibinfo {year} {2000})}\BibitemShut {NoStop}%
\bibitem [{\citenamefont {Liu}\ \emph {et~al.}(2018)\citenamefont {Liu},
  \citenamefont {Lin},\ and\ \citenamefont {Tom{\'a}nek}}]{tomanek}%
  \BibitemOpen
  \bibfield  {author} {\bibinfo {author} {\bibfnamefont {D.}~\bibnamefont
  {Liu}}, \bibinfo {author} {\bibfnamefont {X.}~\bibnamefont {Lin}}, \ and\
  \bibinfo {author} {\bibfnamefont {D.}~\bibnamefont {Tom{\'a}nek}},\ }\href
  {\doibase 10.1021/acs.nanolett.8b01639} {\bibfield  {journal} {\bibinfo
  {journal} {Nano Lett.}\ }\textbf {\bibinfo {volume} {18}},\ \bibinfo {pages}
  {4908} (\bibinfo {year} {2018})}\BibitemShut {NoStop}%
\bibitem [{\citenamefont {H\"anggi}\ \emph {et~al.}(1990)\citenamefont
  {H\"anggi}, \citenamefont {Talkner},\ and\ \citenamefont
  {Borkovec}}]{rmp1990}%
  \BibitemOpen
  \bibfield  {author} {\bibinfo {author} {\bibfnamefont {P.}~\bibnamefont
  {H\"anggi}}, \bibinfo {author} {\bibfnamefont {P.}~\bibnamefont {Talkner}}, \
  and\ \bibinfo {author} {\bibfnamefont {M.}~\bibnamefont {Borkovec}},\ }\href
  {\doibase 10.1103/RevModPhys.62.251} {\bibfield  {journal} {\bibinfo
  {journal} {Rev. Mod. Phys.}\ }\textbf {\bibinfo {volume} {62}},\ \bibinfo
  {pages} {251} (\bibinfo {year} {1990})}\BibitemShut {NoStop}%
\bibitem [{\citenamefont {Pacheco-Sanjuan}\ \emph {et~al.}(2019)\citenamefont
  {Pacheco-Sanjuan}, \citenamefont {Bishop}, \citenamefont {Farmer},
  \citenamefont {Kumar},\ and\ \citenamefont {Barraza-Lopez}}]{elastic}%
  \BibitemOpen
  \bibfield  {author} {\bibinfo {author} {\bibfnamefont {A.}~\bibnamefont
  {Pacheco-Sanjuan}}, \bibinfo {author} {\bibfnamefont {T.~B.}\ \bibnamefont
  {Bishop}}, \bibinfo {author} {\bibfnamefont {E.~E.}\ \bibnamefont {Farmer}},
  \bibinfo {author} {\bibfnamefont {P.}~\bibnamefont {Kumar}}, \ and\ \bibinfo
  {author} {\bibfnamefont {S.}~\bibnamefont {Barraza-Lopez}},\ }\href {\doibase
  10.1103/PhysRevB.99.104108} {\bibfield  {journal} {\bibinfo  {journal} {Phys.
  Rev. B}\ }\textbf {\bibinfo {volume} {99}},\ \bibinfo {pages} {104108}
  (\bibinfo {year} {2019})}\BibitemShut {NoStop}%
\bibitem [{\citenamefont {Littlewood}(1980)}]{Littlewood1}%
  \BibitemOpen
  \bibfield  {author} {\bibinfo {author} {\bibfnamefont {P.~B.}\ \bibnamefont
  {Littlewood}},\ }\href {\doibase 10.1088/0022-3719/13/26/009} {\bibfield
  {journal} {\bibinfo  {journal} {J. Phys. C: Solid State Phys.}\ }\textbf
  {\bibinfo {volume} {13}},\ \bibinfo {pages} {4855} (\bibinfo {year}
  {1980})}\BibitemShut {NoStop}%
\bibitem [{\citenamefont {Ji}\ \emph {et~al.}(2019)\citenamefont {Ji},
  \citenamefont {Cai}, \citenamefont {Paudel}, \citenamefont {Sun},
  \citenamefont {Zhang}, \citenamefont {Han}, \citenamefont {Wei},
  \citenamefont {Zang}, \citenamefont {Gu}, \citenamefont {Zhang},
  \citenamefont {Gao}, \citenamefont {Huyan}, \citenamefont {Guo},
  \citenamefont {Wu}, \citenamefont {Gu}, \citenamefont {Tsymbal},
  \citenamefont {Wang}, \citenamefont {Nie},\ and\ \citenamefont
  {Pan}}]{Nature2019}%
  \BibitemOpen
  \bibfield  {author} {\bibinfo {author} {\bibnamefont {Ji}}, \bibinfo {author}
  {\bibfnamefont {S.}~\bibnamefont {Cai}}, \bibinfo {author} {\bibfnamefont
  {T.~R.}\ \bibnamefont {Paudel}}, \bibinfo {author} {\bibfnamefont
  {H.}~\bibnamefont {Sun}}, \bibinfo {author} {\bibfnamefont {C.}~\bibnamefont
  {Zhang}}, \bibinfo {author} {\bibfnamefont {L.}~\bibnamefont {Han}}, \bibinfo
  {author} {\bibfnamefont {Y.}~\bibnamefont {Wei}}, \bibinfo {author}
  {\bibfnamefont {Y.}~\bibnamefont {Zang}}, \bibinfo {author} {\bibfnamefont
  {M.}~\bibnamefont {Gu}}, \bibinfo {author} {\bibfnamefont {Y.}~\bibnamefont
  {Zhang}}, \bibinfo {author} {\bibfnamefont {W.}~\bibnamefont {Gao}}, \bibinfo
  {author} {\bibfnamefont {H.}~\bibnamefont {Huyan}}, \bibinfo {author}
  {\bibfnamefont {W.}~\bibnamefont {Guo}}, \bibinfo {author} {\bibfnamefont
  {D.}~\bibnamefont {Wu}}, \bibinfo {author} {\bibfnamefont {Z.}~\bibnamefont
  {Gu}}, \bibinfo {author} {\bibfnamefont {E.~Y.}\ \bibnamefont {Tsymbal}},
  \bibinfo {author} {\bibfnamefont {P.}~\bibnamefont {Wang}}, \bibinfo {author}
  {\bibfnamefont {Y.}~\bibnamefont {Nie}}, \ and\ \bibinfo {author}
  {\bibfnamefont {X.}~\bibnamefont {Pan}},\ }\href {\doibase
  10.1038/s41586-019-1255-7} {\bibfield  {journal} {\bibinfo  {journal}
  {Nature}\ }\textbf {\bibinfo {volume} {570}},\ \bibinfo {pages} {87}
  (\bibinfo {year} {2019})}\BibitemShut {NoStop}%
\bibitem [{\citenamefont {Huang}\ \emph {et~al.}(2014)\citenamefont {Huang},
  \citenamefont {Sutter}, \citenamefont {Sadowski}, \citenamefont {Cotlet},
  \citenamefont {Monti}, \citenamefont {Racke}, \citenamefont {Neupane},
  \citenamefont {Wickramaratne}, \citenamefont {Lake}, \citenamefont
  {Parkinson},\ and\ \citenamefont {Sutter}}]{ref18}%
  \BibitemOpen
  \bibfield  {author} {\bibinfo {author} {\bibfnamefont {Y.}~\bibnamefont
  {Huang}}, \bibinfo {author} {\bibfnamefont {E.}~\bibnamefont {Sutter}},
  \bibinfo {author} {\bibfnamefont {J.~T.}\ \bibnamefont {Sadowski}}, \bibinfo
  {author} {\bibfnamefont {M.}~\bibnamefont {Cotlet}}, \bibinfo {author}
  {\bibfnamefont {O.~L.}\ \bibnamefont {Monti}}, \bibinfo {author}
  {\bibfnamefont {D.~A.}\ \bibnamefont {Racke}}, \bibinfo {author}
  {\bibfnamefont {M.~R.}\ \bibnamefont {Neupane}}, \bibinfo {author}
  {\bibfnamefont {D.}~\bibnamefont {Wickramaratne}}, \bibinfo {author}
  {\bibfnamefont {R.~K.}\ \bibnamefont {Lake}}, \bibinfo {author}
  {\bibfnamefont {B.~A.}\ \bibnamefont {Parkinson}}, \ and\ \bibinfo {author}
  {\bibfnamefont {P.}~\bibnamefont {Sutter}},\ }\href {\doibase
  10.1021/nn504481r} {\bibfield  {journal} {\bibinfo  {journal} {ACS Nano}\
  }\textbf {\bibinfo {volume} {8}},\ \bibinfo {pages} {10743} (\bibinfo {year}
  {2014})}\BibitemShut {NoStop}%
\bibitem [{\citenamefont {Lee}\ \emph {et~al.}(2017)\citenamefont {Lee},
  \citenamefont {Shin}, \citenamefont {Ham}, \citenamefont {Lee}, \citenamefont
  {Choi}, \citenamefont {Park},\ and\ \citenamefont {Jeon}}]{SnS2_2}%
  \BibitemOpen
  \bibfield  {author} {\bibinfo {author} {\bibfnamefont {S.}~\bibnamefont
  {Lee}}, \bibinfo {author} {\bibfnamefont {S.}~\bibnamefont {Shin}}, \bibinfo
  {author} {\bibfnamefont {G.}~\bibnamefont {Ham}}, \bibinfo {author}
  {\bibfnamefont {J.}~\bibnamefont {Lee}}, \bibinfo {author} {\bibfnamefont
  {H.}~\bibnamefont {Choi}}, \bibinfo {author} {\bibfnamefont {H.}~\bibnamefont
  {Park}}, \ and\ \bibinfo {author} {\bibfnamefont {H.}~\bibnamefont {Jeon}},\
  }\href {\doibase 10.1063/1.4982068} {\bibfield  {journal} {\bibinfo
  {journal} {AIP Advances}\ }\textbf {\bibinfo {volume} {7}},\ \bibinfo {pages}
  {045307} (\bibinfo {year} {2017})}\BibitemShut {NoStop}%
\bibitem [{\citenamefont {Xu}\ \emph {et~al.}(2019)\citenamefont {Xu},
  \citenamefont {Zhang}, \citenamefont {Jiang}, \citenamefont {Wang},
  \citenamefont {Chen}, \citenamefont {Hu}, \citenamefont {Gong}, \citenamefont
  {Shang}, \citenamefont {Zhang}, \citenamefont {Jiang},\ and\ \citenamefont
  {Chu}}]{SnS2_3}%
  \BibitemOpen
  \bibfield  {author} {\bibinfo {author} {\bibfnamefont {L.}~\bibnamefont
  {Xu}}, \bibinfo {author} {\bibfnamefont {P.}~\bibnamefont {Zhang}}, \bibinfo
  {author} {\bibfnamefont {H.}~\bibnamefont {Jiang}}, \bibinfo {author}
  {\bibfnamefont {X.}~\bibnamefont {Wang}}, \bibinfo {author} {\bibfnamefont
  {F.}~\bibnamefont {Chen}}, \bibinfo {author} {\bibfnamefont {Z.}~\bibnamefont
  {Hu}}, \bibinfo {author} {\bibfnamefont {Y.}~\bibnamefont {Gong}}, \bibinfo
  {author} {\bibfnamefont {L.}~\bibnamefont {Shang}}, \bibinfo {author}
  {\bibfnamefont {J.}~\bibnamefont {Zhang}}, \bibinfo {author} {\bibfnamefont
  {K.}~\bibnamefont {Jiang}}, \ and\ \bibinfo {author} {\bibfnamefont
  {J.}~\bibnamefont {Chu}},\ }\href {\doibase 10.1002/smll.201904116}
  {\bibfield  {journal} {\bibinfo  {journal} {Small}\ }\textbf {\bibinfo
  {volume} {15}},\ \bibinfo {pages} {1904116} (\bibinfo {year}
  {2019})}\BibitemShut {NoStop}%
\bibitem [{\citenamefont {Gonzalez}\ and\ \citenamefont
  {Oleynik}(2016)}]{Oleynik}%
  \BibitemOpen
  \bibfield  {author} {\bibinfo {author} {\bibfnamefont {J.~M.}\ \bibnamefont
  {Gonzalez}}\ and\ \bibinfo {author} {\bibfnamefont {I.~I.}\ \bibnamefont
  {Oleynik}},\ }\href {\doibase 10.1103/PhysRevB.94.125443} {\bibfield
  {journal} {\bibinfo  {journal} {Phys. Rev. B}\ }\textbf {\bibinfo {volume}
  {94}},\ \bibinfo {pages} {125443} (\bibinfo {year} {2016})}\BibitemShut
  {NoStop}%
\bibitem [{\citenamefont {Hahn}(2002)}]{Hahn}%
  \BibitemOpen
  \bibinfo {editor} {\bibfnamefont {T.}~\bibnamefont {Hahn}},\ ed.,\ \href@noop
  {} {\emph {\bibinfo {title} {International Tables for Crystallography Volume
  A: Space-group symmetry}}},\ \bibinfo {edition} {5th}\ ed.\ (\bibinfo
  {publisher} {Springer},\ \bibinfo {address} {Dordrecht},\ \bibinfo {year}
  {2002})\BibitemShut {NoStop}%
\bibitem [{\citenamefont {Perdew}\ \emph {et~al.}(1996)\citenamefont {Perdew},
  \citenamefont {Burke},\ and\ \citenamefont {Ernzerhof}}]{PBE}%
  \BibitemOpen
  \bibfield  {author} {\bibinfo {author} {\bibfnamefont {J.~P.}\ \bibnamefont
  {Perdew}}, \bibinfo {author} {\bibfnamefont {K.}~\bibnamefont {Burke}}, \
  and\ \bibinfo {author} {\bibfnamefont {M.}~\bibnamefont {Ernzerhof}},\ }\href
  {\doibase 10.1103/PhysRevLett.77.3865} {\bibfield  {journal} {\bibinfo
  {journal} {Phys. Rev. Lett.}\ }\textbf {\bibinfo {volume} {77}},\ \bibinfo
  {pages} {3865} (\bibinfo {year} {1996})}\BibitemShut {NoStop}%
\bibitem [{\citenamefont {Poudel}\ \emph {et~al.}(2019)\citenamefont {Poudel},
  \citenamefont {Villanova},\ and\ \citenamefont {Barraza-Lopez}}]{Shiva}%
  \BibitemOpen
  \bibfield  {author} {\bibinfo {author} {\bibfnamefont {S.~P.}\ \bibnamefont
  {Poudel}}, \bibinfo {author} {\bibfnamefont {J.~W.}\ \bibnamefont
  {Villanova}}, \ and\ \bibinfo {author} {\bibfnamefont {S.}~\bibnamefont
  {Barraza-Lopez}},\ }\href {\doibase 10.1103/PhysRevMaterials.3.124004}
  {\bibfield  {journal} {\bibinfo  {journal} {Phys. Rev. Materials}\ }\textbf
  {\bibinfo {volume} {3}},\ \bibinfo {pages} {124004} (\bibinfo {year}
  {2019})}\BibitemShut {NoStop}%
\bibitem [{\citenamefont {Fei}\ \emph {et~al.}(2015)\citenamefont {Fei},
  \citenamefont {Li}, \citenamefont {Li},\ and\ \citenamefont
  {Yang}}]{fei_apl_2015_ges_gese_sns_snse}%
  \BibitemOpen
  \bibfield  {author} {\bibinfo {author} {\bibfnamefont {R.}~\bibnamefont
  {Fei}}, \bibinfo {author} {\bibfnamefont {W.}~\bibnamefont {Li}}, \bibinfo
  {author} {\bibfnamefont {J.}~\bibnamefont {Li}}, \ and\ \bibinfo {author}
  {\bibfnamefont {L.}~\bibnamefont {Yang}},\ }\href {\doibase
  10.1063/1.4934750} {\bibfield  {journal} {\bibinfo  {journal} {Appl. Phys.
  Lett.}\ }\textbf {\bibinfo {volume} {107}},\ \bibinfo {pages} {173104}
  (\bibinfo {year} {2015})}\BibitemShut {NoStop}%
\bibitem [{\citenamefont {Villanova}\ \emph {et~al.}(2020)\citenamefont
  {Villanova}, \citenamefont {Kumar},\ and\ \citenamefont
  {Barraza-Lopez}}]{Villanova2020PRB}%
  \BibitemOpen
  \bibfield  {author} {\bibinfo {author} {\bibfnamefont {J.~W.}\ \bibnamefont
  {Villanova}}, \bibinfo {author} {\bibfnamefont {P.}~\bibnamefont {Kumar}}, \
  and\ \bibinfo {author} {\bibfnamefont {S.}~\bibnamefont {Barraza-Lopez}},\
  }\href {\doibase 10.1103/PhysRevB.101.184101} {\bibfield  {journal} {\bibinfo
   {journal} {Phys. Rev. B}\ }\textbf {\bibinfo {volume} {101}},\ \bibinfo
  {pages} {184101} (\bibinfo {year} {2020})}\BibitemShut {NoStop}%
\bibitem{suppl}%
  \BibitemOpen
See Supplemental Material at ------ for details of calculations leading to the energy path in Figure 4, and for the electronic properties of the SiS monolayer across that path.
\BibitemShut {NoStop}%
\bibitem [{\citenamefont {Noor-A-Alam}\ \emph {et~al.}(2014)\citenamefont
  {Noor-A-Alam}, \citenamefont {Kim},\ and\ \citenamefont {Shin}}]{C3CP53971G}%
  \BibitemOpen
  \bibfield  {author} {\bibinfo {author} {\bibfnamefont {M.}~\bibnamefont
  {Noor-A-Alam}}, \bibinfo {author} {\bibfnamefont {H.~J.}\ \bibnamefont
  {Kim}}, \ and\ \bibinfo {author} {\bibfnamefont {Y.-H.}\ \bibnamefont
  {Shin}},\ }\href {\doibase 10.1039/C3CP53971G} {\bibfield  {journal}
  {\bibinfo  {journal} {Phys. Chem. Chem. Phys.}\ }\textbf {\bibinfo {volume}
  {16}},\ \bibinfo {pages} {6575} (\bibinfo {year} {2014})}\BibitemShut
  {NoStop}%
\bibitem [{\citenamefont {Saito}\ \emph {et~al.}(1998)\citenamefont {Saito},
  \citenamefont {Dresselhaus},\ and\ \citenamefont
  {Dresselhaus}}]{Dresselhaus}%
  \BibitemOpen
  \bibfield  {author} {\bibinfo {author} {\bibfnamefont {R.}~\bibnamefont
  {Saito}}, \bibinfo {author} {\bibfnamefont {G.}~\bibnamefont {Dresselhaus}},
  \ and\ \bibinfo {author} {\bibfnamefont {M.~S.}\ \bibnamefont
  {Dresselhaus}},\ }\href@noop {} {\emph {\bibinfo {title} {Physical Properties
  of Carbon Nanotubes}}},\ \bibinfo {edition} {1st}\ ed.\ (\bibinfo
  {publisher} {World Scientific},\ \bibinfo {address} {Singapore},\ \bibinfo
  {year} {1998})\BibitemShut {NoStop}%
\bibitem [{\citenamefont {Landolt}\ \emph {et~al.}(1987)\citenamefont
  {Landolt}, \citenamefont {B{\"o}rnstein}, \citenamefont {Fischer},
  \citenamefont {Madelung},\ and\ \citenamefont {Deuschle}}]{Landolt}%
  \BibitemOpen
  \bibfield  {author} {\bibinfo {author} {\bibfnamefont {H.}~\bibnamefont
  {Landolt}}, \bibinfo {author} {\bibfnamefont {R.}~\bibnamefont
  {B{\"o}rnstein}}, \bibinfo {author} {\bibfnamefont {H.}~\bibnamefont
  {Fischer}}, \bibinfo {author} {\bibfnamefont {O.}~\bibnamefont {Madelung}}, \
  and\ \bibinfo {author} {\bibfnamefont {G.}~\bibnamefont {Deuschle}},\
  }\href@noop {} {\emph {\bibinfo {title} {Landolt-B{\"o}rnstein: Numerical
  Data and Functional Relationships in Science and Technology}}},\ \bibinfo
  {series} {Numerical Data and Functional Relationships in Science and
  Technology Series}\ No.\ \bibinfo {number} {v. 17}\ (\bibinfo  {publisher}
  {Springer},\ \bibinfo {year} {1987})\BibitemShut {NoStop}%
\bibitem [{\citenamefont {Krukau}\ \emph {et~al.}(2006)\citenamefont {Krukau},
  \citenamefont {Vydrov}, \citenamefont {Izmaylov},\ and\ \citenamefont
  {Scuseria}}]{HSE06}%
  \BibitemOpen
  \bibfield  {author} {\bibinfo {author} {\bibfnamefont {A.~V.}\ \bibnamefont
  {Krukau}}, \bibinfo {author} {\bibfnamefont {O.~A.}\ \bibnamefont {Vydrov}},
  \bibinfo {author} {\bibfnamefont {A.~F.}\ \bibnamefont {Izmaylov}}, \ and\
  \bibinfo {author} {\bibfnamefont {G.~E.}\ \bibnamefont {Scuseria}},\ }\href
  {\doibase 10.1063/1.2404663} {\bibfield  {journal} {\bibinfo  {journal} {J.
  Chem. Phys.}\ }\textbf {\bibinfo {volume} {125}},\ \bibinfo {pages} {224106}
  (\bibinfo {year} {2006})}\BibitemShut {NoStop}%
\bibitem [{\citenamefont {Thomas}\ \emph {et~al.}(2016)\citenamefont {Thomas},
  \citenamefont {Ajith},\ and\ \citenamefont {Valsakumar}}]{c66}%
  \BibitemOpen
  \bibfield  {author} {\bibinfo {author} {\bibfnamefont {S.}~\bibnamefont
  {Thomas}}, \bibinfo {author} {\bibfnamefont {K.~M.}\ \bibnamefont {Ajith}}, \
  and\ \bibinfo {author} {\bibfnamefont {M.~C.}\ \bibnamefont {Valsakumar}},\
  }\href {\doibase 10.1088/0953-8984/28/29/295302} {\bibfield  {journal}
  {\bibinfo  {journal} {J. Phys.: Condens. Matter}\ }\textbf {\bibinfo {volume}
  {28}},\ \bibinfo {pages} {295302} (\bibinfo {year} {2016})}\BibitemShut
  {NoStop}%
\end{thebibliography}

%

\end{document}